\documentclass[lettersize,journal]{IEEEtran}

\usepackage{tabu}                      
\usepackage{booktabs}                  
\usepackage{lipsum}                    
\usepackage{mwe}                       

\usepackage{mathptmx}                  

\usepackage{amsmath,amsfonts}
\usepackage{algorithmic}
\usepackage{algorithm}
\usepackage{array}
\usepackage[caption=false,font=normalsize,labelfont=sf,textfont=sf]{subfig}
\usepackage{textcomp}
\usepackage{stfloats}
\usepackage{url}
\usepackage{verbatim}
\usepackage{graphicx}
\usepackage{cite}

\usepackage{enumitem}
\usepackage{fontawesome5}
\usepackage{xcolor}
\usepackage{tabularx}
\usepackage{colortbl}
\usepackage{tikz}
\usepackage{mdframed}
\newcommand{\app}{ProvenanceLens\xspace}

\newcommand{\user}[1]{\textcolor{purple}{$P_{#1}$}}

\newcommand{\paragraphHeadingSpace}{\vspace{4px}}
\newcommand{\bpstart}[1]{
{\textbf{#1}}%
}

\definecolor{cmaroon}{RGB}{192,0,0}
\definecolor{cbrown}{RGB}{197,90,17}
\definecolor{cpurple}{RGB}{112,48,159}
\definecolor{cgreen}{RGB}{84,130,53}
\definecolor{cblue}{RGB}{0,112,192}
\definecolor{clightgray}{RGB}{230, 230, 230}

\definecolor{cglyphlightblue}{RGB}{207,224,242}
\definecolor{cglyphblue}{RGB}{91,163,207}
\definecolor{cglyphdarkblue}{RGB}{9,74,144}

\newcommand{\orangecircleglyph}[2][orange]{%
  \tikz[baseline=-0.75ex]{
    \node[shape=circle,draw,#1,fill=#1,inner sep=0pt, minimum size=8pt, text=white, font=\footnotesize] (char) {#2};}
}

\newcommand{\lightbluesquareglyph}[2][cglyphlightblue]{%
  \tikz[baseline=-0.75ex]{
    \node[shape=rectangle,draw,#1,fill=#1,inner sep=0pt, text=white, font=\footnotesize, minimum width=7pt, minimum height=7pt] (char) {#2};}
}

\newcommand{\bluesquareglyph}[2][cglyphblue]{%
  \tikz[baseline=-0.75ex]{
    \node[shape=rectangle,draw,#1,fill=#1,inner sep=0pt, text=white, font=\footnotesize, minimum width=7pt, minimum height=7pt] (char) {#2};}
}

\newcommand{\darkbluesquareglyph}[2][cglyphdarkblue]{%
  \tikz[baseline=-0.75ex]{
    \node[shape=rectangle,draw,#1,fill=#1,inner sep=0pt, text=white, font=\footnotesize, minimum width=7pt, minimum height=7pt] (char) {#2};}
}

\newcommand{\lightbluecircleglyph}[2][cglyphlightblue]{%
  \tikz[baseline=-0.75ex]{
    \node[shape=circle,draw,#1,fill=#1,inner sep=0pt, minimum size=8pt, text=white, font=\footnotesize] (char) {#2};}
}

\newcommand{\bluecircleglyph}[2][cglyphblue]{%
  \tikz[baseline=-0.75ex]{
    \node[shape=circle,draw,#1,fill=#1,inner sep=0pt, minimum size=8pt, text=white, font=\footnotesize] (char) {#2};}
}

\newcommand{\darkbluecircleglyph}[2][cglyphdarkblue]{%
  \tikz[baseline=-0.75ex]{
    \node[shape=circle,draw,#1,fill=#1,inner sep=0pt, minimum size=8pt, text=white, font=\footnotesize] (char) {#2};}
}

\newcommand{\bluethickstrokecircleglyph}[2][cblue]{%
  \tikz[baseline=-0.75ex]{
    \node[shape=circle,draw,#1,inner sep=0pt, minimum size=8pt, text=black, line width=1.25pt, font=\footnotesize] (char) {#2};}
}

\newcommand{\bluethinstrokecircleglyph}[2][cblue]{%
  \tikz[baseline=-0.75ex]{
    \node[shape=circle,draw,#1,inner sep=0pt, minimum size=8pt, text=black, line width=0.5pt, font=\footnotesize] (char) {#2};}
}

\newmdenv[
  leftmargin=0pt,
  rightmargin=0pt,
  innerleftmargin=4pt,
  innerrightmargin=4pt,
  innertopmargin=4pt,
  innerbottommargin=4pt,
  backgroundcolor=clightgray,
  linecolor=clightgray
]{highlight}

\usepackage{setspace} 
\begin{document}

\title{Utilizing Provenance as an Attribute for Visual Data Analysis: A Design Probe with ProvenanceLens}

\author{
Arpit Narechania, Shunan Guo, Eunyee Koh, Alex Endert, Jane Hoffswell
\thanks{Arpit Narechania is with The Hong Kong University of Science and Technology (arpit@ust.hk); this work was done in part when he was with The Georgia Institute of Technology. Alex Endert is with The Georgia Institute of Technology (endert@gatech.edu). Shunan Guo, Eunyee Koh, and Jane Hoffswell are with Adobe Research~(\{sguo, eunyee, jhoffs\}@adobe.com).}
}




\maketitle

\begin{abstract}
Analytic provenance can be visually encoded to~help users track their ongoing analysis trajectories, recall past interactions, and inform new analytic directions. 
Despite its significance, provenance is often hardwired into analytics systems, affording limited user control and opportunities for self-reflection. 
We thus propose modeling \textbf{provenance as an attribute} that is available to users during analysis.
We demonstrate this concept by modeling two provenance attributes that track the \emph{recency} and \emph{frequency} of user interactions with data.
We integrate these attributes into a visual data analysis system prototype, \textbf{ProvenanceLens}, wherein users can visualize their interaction recency and frequency by mapping them to encoding channels~(e.g., color, size) or applying data transformations~(e.g., filter, sort).
Using ProvenanceLens as a design probe, we conduct an exploratory study with sixteen users to investigate how these provenance-tracking affordances are utilized for both decision-making and self-reflection. 
We find that users can accurately and confidently answer questions about their analysis, and we show that mismatches between the user's mental model and the provenance encodings can be surprising, thereby prompting useful self-reflection. 
We also report on the user strategies surrounding these affordances, and reflect on their intuitiveness and effectiveness in representing provenance.
\end{abstract}


\section{Introduction}
Analytic provenance records the history of analytical actions, showing how data was obtained, transformed, and analyzed. For data visualization, analytic provenance also tracks how users interact with visualizations as a representation of their reasoning process~\cite{north2011analytic}.
A frequent use of provenance is to help users recall steps taken during analysis~\cite{ragan2015characterizing}.
While effective for forensic purposes, other tools have explored how to show provenance to users during analysis.
For example, existing systems~\cite{narechania2022lumos,feng2017hindsight,willett2007scented} leave visual traces of the user's interactions to encourage them to pause and reflect on their behavior, potentially influencing subsequent analysis.
However, analytic~questions such as \emph{``How many data points has the user interacted with so far?''} or \emph{``Which were the first attributes that the user interacted with?''} are often only answerable post-analysis (after analyzing the provenance logs). 
In the moment, it is impractical for the user to manually count the interaction traces one-by-one or outright remember their interaction history in detail, which makes such iterative reflection during analysis difficult~\cite{miller1994magical, liu2014effects, stitz2019knowledgepearls}.
Thus, our first research question asks: \emph{``How can we make (such) provenance information available to the user during visual data analysis?''}
While prior work has already explored modeling~\cite{zhou2021modeling} and presentation~\cite{narechania2022lumos} of analytic provenance, this work aims to \emph{standardize} these aspects for consistent and flexible access during analysis.

For modeling provenance, we introduce two \textbf{provenance attributes}, \emph{frequency} and \emph{recency}. 
These attributes track how often and how recently a user interacts with each data attribute and record, assigning a score that is standardized to a common scale from 0.0 (no interaction) to 1.0 (most frequent or most recent interaction, respectively).
By modeling the \textit{frequency} of interactions this way, analysts can quickly identify unexplored data, or ones that they keep coming back to for subsequent decision-making. \textit{Recency} on the other hand lets users track when data are interacted with and review how their analytic behavior evolves over time.

For presenting the modeled provenance to users, we propose to dynamically map provenance attributes to one or more visual encoding channels.
For example, users can map \emph{frequency} to the color and/or size channels, creating a visualization where data points are colored and/or sized based on their frequency scores, providing a summary of past interactions and influencing subsequent ones.
This flexibility contrasts with prior work, where provenance is often \emph{permanently mapped} to specific encodings (e.g., color~\cite{narechania2022lumos,narechania2025provenancewidgets}, size~\cite{narechania2025provenancewidgets}, opacity~\cite{feng2017hindsight}, or stroke width~\cite{feng2017hindsight}), preventing users from temporarily disabling them or mapping other information to those channels.

In addition to visual encodings, we also propose sorting and filtering (i.e., transforming data) by provenance attributes to help users strategize about subsequent interactions, such as organizing the data, e.g., ``rank data attributes by their frequency'' (sort) or reducing the search space, e.g., ``show top five most recently interacted points'' (filter).
Collectively, we standardize provenance visualization for data attributes and records while also affording more user flexibility to encode, filter, and sort provenance, or combinations thereof.

Such modeling and presentation of provenance information motivates our second, central research question, \emph{``How do people use the provenance attributes during visual data analysis?''}
For example, do users prefer mapping provenance to color, size, both color and size, or something different, like an axis?
If not visual encodings, do users prefer interacting with provenance via data transformations such as sorting or filtering?
Lastly, do these user preferences change for specific tasks, e.g., while reviewing someone else's analysis history versus doing one's own analysis?

To study the utility and usage characteristics of provenance attributes during visual analysis, we developed a system prototype, \textbf{\app}, that allows users to map provenance to visual encodings and data transformations; 
users can now interact with provenance in the same way as regular~attributes.

Using \app as a design probe, we conducted an \textbf{exploratory study} with sixteen users to investigate how they utilized provenance attributes for a decision-making task. 
Specifically, we tasked users to (1)~\emph{review} and \emph{answer} questions about another user's provenance that is preloaded into the system and then (2)~\emph{track} and \emph{answer} questions about their own provenance.
Our users also self-reported their confidence and surprise levels while answering these questions.

We found that provenance attributes in \app can help users accurately and confidently answer questions about another user's as well as their own analysis. 
Users were often surprised
and uncovered unexpected insights and analytical behaviors that they may have misremembered, facilitating self-reflection.
In addition to user performance, we report on user preferences and strategies for utilizing available affordances, assessing their intuitiveness and effectiveness to track and recall analytic provenance.
For instance, we found that \emph{x}, \emph{y}, and \emph{fill} were the most preferred encodings, whereas, \emph{shape} and \emph{stroke} were either rarely used or not used at all.
We conclude by discussing design implications for future provenance-based systems and studies.
Our primary contributions include:
\vspace{4px}
\begin{enumerate}[nosep]
    \item Modeling and presentation of the \emph{frequency} and \emph{recency} of \textbf{analytic provenance as attributes (i.e., ``provenance attributes'')} during visual data analysis.
    \item Findings from an exploratory study with sixteen users
    on \textbf{how they use provenance attributes in a visual analysis system} to accurately and confidently answer questions about their analysis, sometimes causing surprise.
\end{enumerate}
\label{section:introduction}

\section{Related Work}
\label{section:relatedwork}
\subsection{Analytic Provenance: Tracking, Modeling, and Visualizing}
\label{sec:analyticprovenance}

Our memory has a finite capacity to track and remember our past interactions with data, which affects exploration~\mbox{\cite{miller1994magical, liu2014effects}}.
Analyzing prior data interactions in a visualization is a form of analytic provenance~\cite{north2011analytic, ragan2015characterizing} that is often used to infer one's analysis process.
Studying such analytic provenance has a well-documented history in visualization research.
Heer~et~al. have summarized the design space for displaying interaction histories~\cite{heer2008graphical}.
Xu~et~al.~\cite{xu2020survey}'s survey notes that provenance is broadly used to support sensemaking~\cite{perry2009supporting,nguyen2016sensemap} and decision-making~\cite{madanagopal2019analytic} with data, evaluate the usefulness of visualization systems~\cite{bylinskii2017learning, gomez2012modeling}, design adaptive systems~\cite{ceneda2016characterizing, walch2019lightguider}, improve the performance of machine learning models~\cite{endert2012semanticinteraction}, replay or replicate analysis sessions~\cite{bavoil2005vistrails, shrinivasan2009connecting}, and automatically generate summary reports of an analysis session~\cite{chen2010click2annotate, gratzl2016visual}.
Provenance also increases awareness of analytic behavior~\cite{narechania2022lumos},
mitigates human biases~\cite{wall2022lrg}, increases unique data discoveries~\cite{feng2017hindsight, willett2007scented}, and affects confidence levels~\cite{block2023influence} during analysis.

Today, provenance tracking occurs in various contexts including analysis tools~\cite{north2011analytic}, code editors~\cite{footstepsvscode}, computational notebooks~\cite{gadhave2024persist, eckelt2024loops}, workflow modeling systems~\cite{bavoil2005vistrails,revisit2023ding}, collaborative environments~\cite{ellkvist2008using,sarvghad2015exploiting,badam2017supporting}, websites~\cite{googleanalytics, hotjar}, and video games~\cite{drachen2015behavioral, kohwalter2017capturing}.
There are several logging frameworks that help capture this provenance~\cite{callahan2006vistrails,aigner2013evalbench,okoe2015graphunit,cutler2020trrack}.
Relevant to this work on visual data analysis, provenance has been modeled from user interactions (via mouse~\cite{arroyo2006usability} or gaze~\cite{nielsen2010eyetracking} tracking) with UI controls~\cite{narechania2025provenancewidgets, baudisch2006phosphor, hill2003awareness, angelini2020crosswidgets}, visualizations as a whole~\cite{gutwin2002traces,boy2015storytelling,feng2017hindsight,stitz2019knowledgepearls} or with specific elements and regions within them (e.g., marks)~\cite{ragan2015characterizing,north2011analytic,feng2017hindsight}, including attributes and records~\cite{narechania2022lumos, wall2022lrg}.
Several metrics have been proposed that quantify analytic focus~\cite{zhou2021modeling}, exploration behavior~\cite{feng2019patterns} and subsequent interaction predictions~\cite{ottley2019follow}, 
differences between a user's subset selections and the data~\cite{gotz2016adaptive}, and differences between a user's interactions and a baseline~\cite{wall2017warning}.
Our work extends previous efforts to model provenance from mouse-based interactions with data by proposing a more standardized approach (i.e., the concept of ``provenance attributes'') that facilitates flexible, real-time user access to this provenance during analysis.

Lastly, provenance visualizations often involve encoding provenance on or near the object of interaction, e.g., highlighting previously visited or interacted 
data attributes and records~\cite{narechania2022lumos}, 
visualizations as a whole~\cite{feng2017hindsight,sarvghad2015exploiting,badam2017supporting} or specific regions within them~\cite{skopik2005improving}, 
regions and hyperlinks on a webpage~\cite{kaasten2002people,hotjar}, 
options and ranges in UI controls~\mbox{\cite{willett2007scented, narechania2025provenancewidgets},}
lines of code in a code editor~\cite{footstepsvscode}, 
or a document's authorship and readership history~\cite{hill1992edit,alexander2009revisiting}, among others.
Other provenance visualizations are separately visualized in an external view or application, e.g., as a graph~\cite{bavoil2005vistrails, cutler2020trrack, oliveira2017framework, kohwalter2016prov, chen2012visualization}.
Relevant to our work, Lumos~\cite{narechania2022lumos} tracks the frequency of user interactions with data and leaves behind visual traces (using color) to increase user awareness of analytic behaviors.
Our work extends Lumos~\cite{narechania2022lumos} to track not only the frequency but also the recency of interactions with data, and to present provenance via not only color but also a variety of visual encodings (e.g., size, shape, tooltip) and data transformations.

\subsection{Direct Manipulation}
\label{subsection:directmanipulation}

Direct manipulation is an interaction technique wherein objects of interest in the UI are visible and users can manipulate them through physical, reversible, and incremental actions and receive immediate feedback~\cite{shneiderman1983direct}. 
This technique has been effectively used to enhance user interaction and comprehension during visual data analysis~\cite{wolter2009direct, kondo2014dimpvis, lex2014upset}. 
For instance, Tableau~\cite{tableau} allows users to perform drag-and-drop operations to specify data visualizations and apply data transformations. 
Wrangler~\cite{kandel2011wrangler} utilizes direct manipulation to help users author expressive transformations while simplifying specification and minimizing manual repetition.
Relevant to our work, DataPilot~\cite{narechania2023datapilot} helps users to navigate large, unfamiliar datasets by enabling them to sort and filter data attributes and records based on their quality and usage characteristics~\cite{narechania2023datacockpit}.
We similarly enable users to sort and filter data but based on provenance (i.e., recency, frequency of their interactions) during analysis.

\section{Utilizing Provenance as an Attribute\\ during Visual Data Analysis}
\label{section:provenanceattributes}
\begin{figure}[t]
    \centering
    \includegraphics[width=\linewidth]{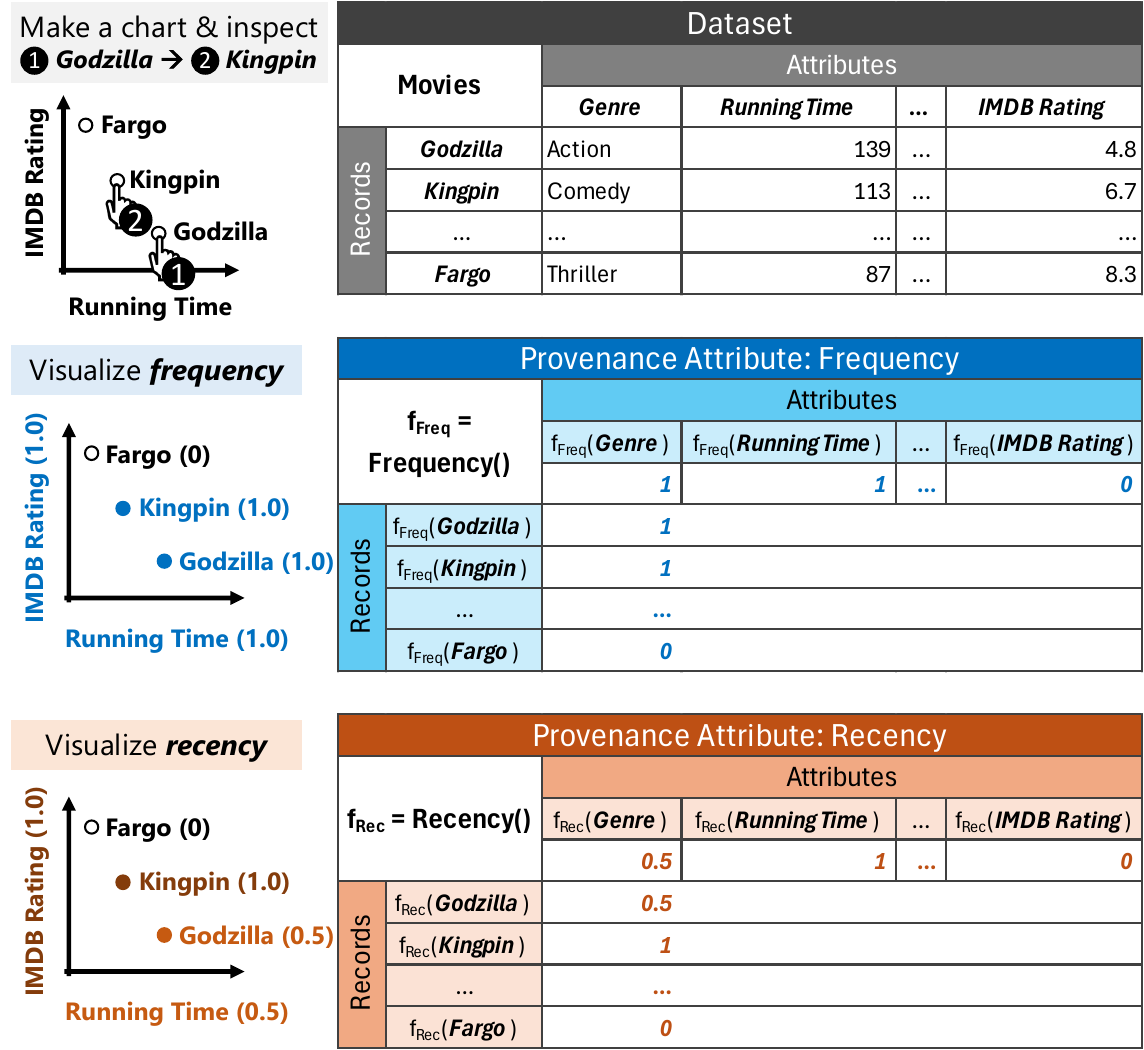}
    \caption{Illustration of two provenance attributes, \textcolor{cblue}{frequency} and \textcolor{cbrown}{recency}, modeled for each attribute and record of a dataset about movies, and scored on a scale from 0 (low) to 1 (high).
    Consider a user creates a scatterplot visualization of \textbf{(1)~Running~Time}~$\times$~\textbf{(2)~IMDB~Rating} and then clicks two datapoints one after another \textbf{(1) Godzilla $\rightarrow$ (2) Kingpin}, indicating interactions with two attributes and two records.
    Regarding data attributes, \textbf{IMDB~Rating} and \textbf{Running~Time} both receive a \emph{\textcolor{cblue}{frequency}} score of \textcolor{cblue}{1.0} (each interacted once, hence maximum score), while other attributes score \textcolor{cblue}{0.0}; for \emph{\textcolor{cbrown}{recency}}, \textbf{IMDB~Rating} (most recently interacted) scores \textcolor{cbrown}{1.0} and \textbf{Running~Time} scores \textcolor{cbrown}{0.5} (user mapped \emph{x axis} before \emph{y axis}), while other attributes score \textcolor{cbrown}{0.0}.
    Likewise, regarding data records, \textbf{Godzilla} and \textbf{Kingpin} both score \textcolor{cblue}{1.0} on \emph{\textcolor{cblue}{frequency}}; but for \emph{\textcolor{cbrown}{recency}}, \textbf{Kingpin} (most recently interacted) scores \textcolor{cbrown}{1.0} and \textbf{Godzilla} scores \textcolor{cbrown}{0.5}, while other records score \textcolor{cbrown}{0.0}. 
    These scores are derived by evenly spacing the interactions between 0 and 1, based on their count and order of occurrence in the interaction history.}
    \label{fig:concept}
\end{figure}

In this section, we describe how we (A) track user interactions with data attributes and records to (B) model provenance attributes and subsequently (C) visualize them during analysis.

\subsection{Tracking Provenance: Which User Interactions to Log?}

In a typical visual data analysis system, users analyze~their dataset in various ways: they can inspect an attribute's summary statistics~(e.g., via distribution plots), examine individual data records~(e.g., from a data table), apply data transformations~(e.g., filter and sort), and create visualizations~(e.g.,~by mapping attributes to visual encodings).
To achieve our overarching goal of tracking, modeling, and visualizing provenance during analysis, we track a subset of relevant user interactions and map them to an individual data attribute or record.
In this work, we track interactions with a data \textit{attribute} when a user inspects its summary profile, maps it to a visual encoding, or uses it to filter and sort data records; we track interactions with a data \textit{record} when a user hovers on a visualization mark (e.g., a point in a scatterplot) or a row in the data table.

\subsection{Modeling Provenance Attributes: Frequency, Recency}
\label{subsection:modeling}

After tracking user interactions, accurately modeling provenance attributes is crucial for understanding analytic behaviors. To model provenance in our work, we generally follow the methodology employed by Lumos~\cite{narechania2022lumos}. For unit visualizations, we map one interaction with an attribute or record as $+1$ unit of interaction. 
For aggregate visualizations that show a single value computed from multiple data records (e.g., a bar in a bar chart with an aggregation function such as \emph{mean} or \emph{sum}), we map one interaction with an aggregated entity as $+1/N$ units of interaction for each of the $N$ data records that form the hovered entity. 
For example, consider a bar chart showing average \emph{IMDB Rating} for different movie \emph{Genre}s; if the user hovers on the bar \emph{Genre}=``Action'' that represents five action movies (whose mean is encoded as the bar's height), we log each action movie as having $+0.2$ units of interaction.

We also log the timestamp (as milliseconds since epoch)~of each interaction. For unit visualizations, we simply map the interaction timestamp to the corresponding attribute or record. For aggregate visualizations, we log the same interaction timestamp for all data records that form the aggregated entity (i.e., many records will have the same interaction timestamp).

From the interaction units and timestamps, we compute~two metrics that we refer to as \textbf{provenance attributes}: \emph{frequency} and \emph{recency}.
We chose to focus on these metrics as they are both relevant to provenance tracking and have been commonly used in visualization research and practice~\cite{narechania2025provenancewidgets, cutler2020trrack,narechania2022lumos,wall2022lrg,zhou2021modeling}.
Figure~\ref{fig:concept} illustrates sample \emph{frequency} and \emph{recency} computations, which are described in the following paragraphs.

\paragraphHeadingSpace\bpstart{Frequency.} This provenance attribute computes a frequency score \( f_{x} \) normalized from zero to one, for each data attribute:

\vspace{-0.5em}
\[
f_{x} = \frac{n_{x}}{\max_{i=1}^{N} n_i}
\]
\vspace{-0.75em}

\noindent where \( n_{x} \) is the total number of interactions with a data attribute~\( x \) and \( \max_{i=1}^{N} n_i \) is the maximum number of interactions from among all~\( N \) data attributes. A score of zero implies no interactions (or zero focus) and a score of one implies the most number of interactions (or maximum focus). Like data attributes, we also compute \( f_{x} \) for data records.

\paragraphHeadingSpace\bpstart{Recency.} This provenance attribute computes a recency score \( r_{x} \) normalized from zero to one, for each data attribute:

\vspace{-0.5em}
\[
r_{x} = \frac{\text{rank}_N(\text{max}(t_{x}))}{\sum_{i=1}^{N} n_i}
\]
\vspace{-0.75em}

\noindent where \( \text{max}(t_{x}) \) is the timestamp of the most recent interaction with a data attribute~\( x \), \( \text{rank}_N(\text{max}(t_{x})) \) is its serial order in the overall sequence of interactions across all~\( N \) data attributes (i.e., the entire analysis history), and \( \sum_{i=1}^{N} n_i \) is the total number of interactions across all~\( N \) data attributes. 
A score of zero implies no interactions (or zero focus) and a score of one corresponds to the most recent interaction (or focus). Like data attributes, we also compute \( r_{x} \) for data records.

\subsection{Visualizing \& Interacting with Provenance during Analysis}
\label{subsection:presenting}

\begin{figure*}[ht]
    \centering
    \includegraphics[width=\linewidth]{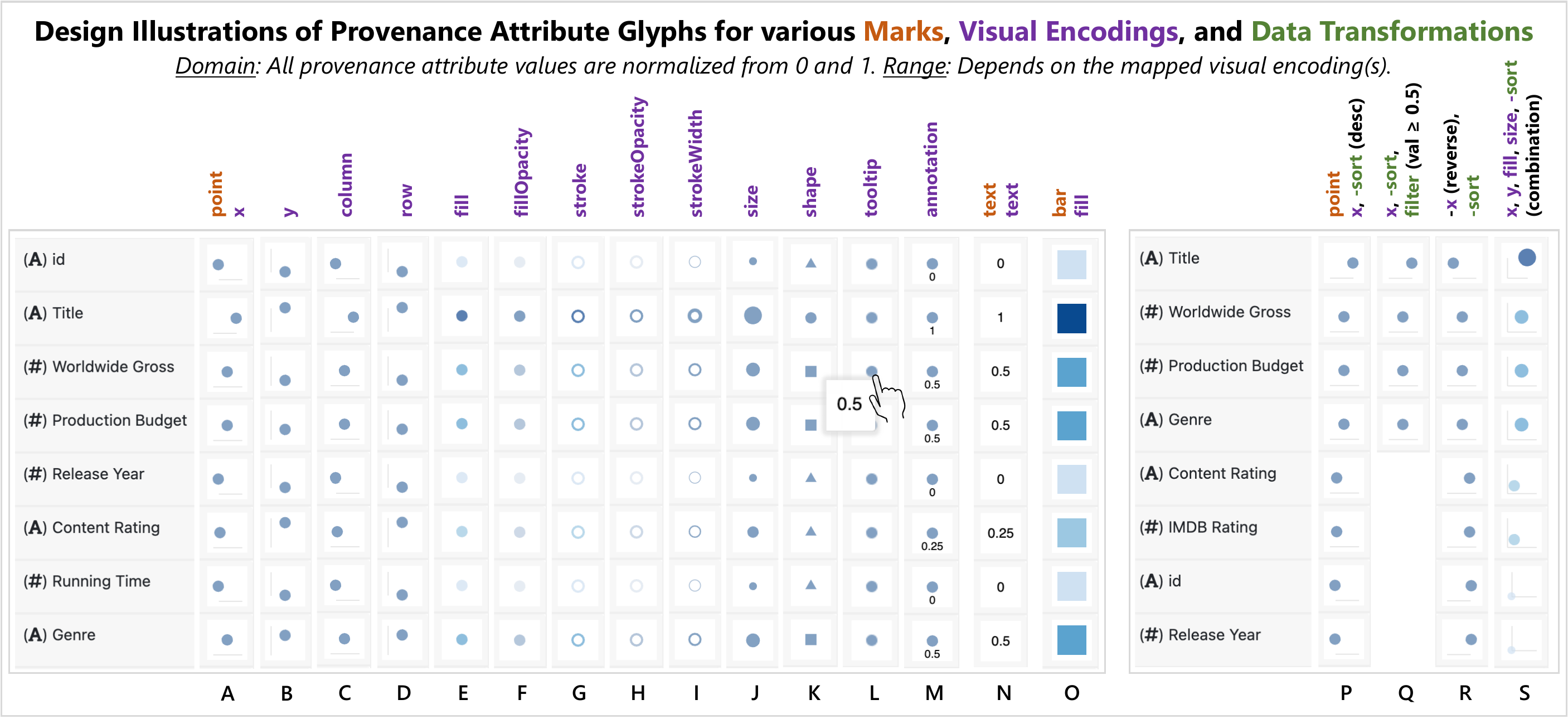}
    \caption{
        Design illustrations of \textbf{\textcolor{cblue}{provenance attribute glyphs}} (\textbf{A}--\textbf{S}) for data attributes (or records) across different \textbf{\textcolor{cbrown}{marks}} (\textbf{point}, \textbf{text}, \textbf{bar}), \textbf{\textcolor{cpurple}{visual encodings}} (\textbf{x}, \textbf{y}, \textbf{column}, \textbf{row}, \textbf{fill}, \textbf{fillOpacity}, \textbf{stroke}, \textbf{strokeOpacity}, \textbf{strokeWidth}, \textbf{size}, \textbf{shape}, \textbf{tooltip}, \textbf{annotation}, \textbf{text}), and \textbf{\textcolor{cgreen}{data transformations}} (\textbf{sort}, \textbf{filter}), including alternate configurations (e.g., \textbf{--x} where the range is \textbf{reverse}d or \textbf{desc}ending sort order) and combinations (e.g., \textbf{x} + \textbf{y} + \textbf{fill} + \textbf{size} + \textbf{sort}).
        For instance, for \textbf{\textcolor{cbrown}{mark}=\textcolor{cbrown}{bar}} and \textbf{\textcolor{cpurple}{encoding}=\textcolor{cpurple}{fill}} (\textbf{O}): ``Title''~\darkbluesquareglyph{}\kern0px has the largest value (darkest bar) followed by ``Worldwide Gross''~\bluesquareglyph{}\kern-2px, ``Production Budget''~\bluesquareglyph{}\kern-2px, and ``Genre''~\bluesquareglyph{}\kern-2px; ``id''~\lightbluesquareglyph{}\kern-2px, ``Release Year''~\lightbluesquareglyph{}\kern-2px, and ``Running Time''~\lightbluesquareglyph{} have the smallest values (lightest bars). Notice the change to the attribute sort order for the right side of the figure (\textbf{P}--\textbf{S}), compared to the unsorted attributes on the left (\textbf{A}--\textbf{O}).}
    \label{fig:provenance-glyphs}
\end{figure*}

Our main goals were to enable users to access the provenance of specific attributes or records and also obtain a visual provenance overview of the entire dataset during analysis. 
In response, we designed small glyphs called \textbf{``provenance attribute glyphs''}, which comprise a mark type \mbox{(e.g., \emph{point}~\bluethinstrokecircleglyph{}\kern-3px)} and one or more visual encodings \mbox{(e.g., \emph{fill}\kern2px\bluecircleglyph{}\kern-3px)} that encode~the provenance attribute values (e.g., \lightbluecircleglyph{}\kern-2px\bluecircleglyph{}\kern-2px\darkbluecircleglyph{}where darker glyphs imply higher values).
These glyphs can represent data records within visualizations (e.g., points in a scatterplot), can be displayed alongside data attributes (e.g., in the attribute panel), and can integrate well with sorting and filtering operations. 
These glyphs can reveal interesting provenance patterns, e.g., if there are more dark than light glyphs (or vice versa), then the user has interacted disproportionately.
Figure~\ref{fig:provenance-glyphs} illustrates the supported designs of provenance attribute glyphs for data attributes, described in detail below.

\paragraphHeadingSpace\subsubsection{Marks and Visual Encodings}

We cover three mark types: \emph{point}~\bluethinstrokecircleglyph{}\kern-2px, \emph{bar}~\bluesquareglyph{}\kern-2px, and \emph{text}~(0.5).
Other mark types such as \emph{line} and \emph{area} require at least two values (a start and an end), which make them unsuitable for encoding, and hence are not considered.
Each mark type encodes a single value across one or more visual encodings, described next.

We cover thirteen encodings~\cite{satyanarayan2016vega}: \emph{x}, \emph{y}, \emph{column}, \emph{row}, \emph{fill}, \emph{fillOpacity}, \emph{stroke}, \emph{strokeOpacity}, \emph{strokeWidth}, \emph{size}, \emph{shape}, \emph{tooltip}, \emph{text}, and \emph{annotation}; \emph{annotation} is a special encoding that adds an extra text mark displaying the encoded value next to the visualization mark, unlike the \emph{text} encoding, that only displays the text (as the visual mark itself).

For instance, Figure~\ref{fig:provenance-glyphs}\textbf{O}~(\textbf{\textcolor{cbrown}{mark}=\textcolor{cbrown}{bar}}, \textbf{\textcolor{cpurple}{encoding}=\textcolor{cpurple}{fill}}) shows that ``Title'' has the largest value (darkest bar\kern2px\mbox{\darkbluesquareglyph{}\kern-3px),} whereas ``id'', ``Release Year'', and ``Running Time'' have the smallest values (lightest bar~\lightbluesquareglyph{}\kern-2px).
Similarly, Figure~\ref{fig:provenance-glyphs}\textbf{I}~(\textbf{\textcolor{cbrown}{mark}=\textcolor{cbrown}{point}} and \textbf{\textcolor{cpurple}{encoding}=\textcolor{cpurple}{strokeWidth}}) shows the glyph\kern2px\bluethickstrokecircleglyph{}\kern0px for the largest value (thick stroke) and~\bluethinstrokecircleglyph{}for the smallest value (thin stroke).

\paragraphHeadingSpace\subsubsection{Data Transformations} 
In addition to visualizing provenance attributes (marks and encodings), we also enable users to transform (filter and sort) their data by the provenance attributes.
This functionality was inspired by DataPilot~\cite{narechania2023datapilot}, which similarly lets users sort and filter their data attributes and records based on their quality and usage characteristics. 

\paragraphHeadingSpace\bpstart{Sort} displays the data attributes or records in order of the encoded provenance attributes. For example, Figure~\ref{fig:provenance-glyphs}\textbf{P} shows an active descending \textcolor{cgreen}{\textbf{--sort}} by \emph{frequency},
visualized as \textbf{\textcolor{cbrown}{mark}=\textcolor{cbrown}{point}} and \textbf{\textcolor{cpurple}{encoding}=\textcolor{cpurple}{x}},
illustrating that ``Title'' had the most interactions followed by ``Worldwide Gross'', ``Production Budget'', and so on. Notice the change to the attribute sort order for the right side of Figure~\ref{fig:provenance-glyphs} (\textbf{P}--\textbf{S}), compared to the unsorted attributes on the left (\textbf{A}--\textbf{O}).

\paragraphHeadingSpace\bpstart{Filter} displays a subset of data attributes or records based on the criteria provided by the user (e.g., the encoded entity values). For example, Figure~\ref{fig:provenance-glyphs}\textbf{Q} shows an active \textcolor{cgreen}{\textbf{filter}} for \emph{frequency} greater than or equal to 0.5, 
emphasized by \textbf{\textcolor{cbrown}{mark}=\textcolor{cbrown}{point}}, \textbf{\textcolor{cpurple}{encoding}=\textcolor{cpurple}{x}}, and a descending \textcolor{cgreen}{\textbf{--sort}} (also by \textit{frequency}),
resulting in four attributes that match the filter (``Title'', ``Worldwide Gross'', ``Production Budget'', ``Genre'').

\paragraphHeadingSpace\subsubsection{Configurations and Combinations} 
We enable various configurations and combinations of marks, visual encodings, and data transformations, affording user agency and control, and promoting accessibility during analysis.

\paragraphHeadingSpace\bpstart{Configurations.} Users can configure the range of the encoding scale (e.g., for \emph{size}, high values map to bigger glyphs or vice versa) or sort directions (i.e., ascending or descending).
For example, all data attributes to the right side of Figure~\ref{fig:provenance-glyphs} (\textbf{P}--\textbf{S}) are \textcolor{cgreen}{\textbf{--sort}}ed by \emph{frequency} in the descending order (which means attributes at the top have been used more often in the interface). However, \textbf{P} maps the glyphs to \textcolor{cpurple}{\textbf{x}} whereas \textbf{R} maps them to \textcolor{cpurple}{\textbf{--x}} (reverse).
Such configurability can support different user preferences and use cases, e.g., visited points could become smaller to nudge the user to interact with other points (increase coverage); alternatively, they could become bigger to help the user quickly spot them.

\paragraphHeadingSpace\bpstart{Combinations.} We do not limit the user to one configuration at a time. Users can simultaneously utilize one or more visual encoding assignments, as well as filter and sort criteria to realize a wide variety of glyph combinations. For example, Figure~\ref{fig:provenance-glyphs}\textbf{S} shows a combination of \emph{x}, \emph{y}, \emph{fill}, \emph{size}, and \emph{sort}.
Combinations can help reinforce certain interaction patterns and also help tools ship with smart defaults to serve a wider audience, e.g., using both \emph{fill} and \emph{size} can support both colorblind and non-colorblind users~\cite{franconeri2021scienceviscom}, similar to how figures in articles are often colored but also hatched.

\section{Evaluation: Exploratory User Study}
\label{section:evaluation}
To study the utility and usage characteristics for provenance attributes during visual data analysis, we designed a decision-making task that involves reviewing and answering questions about both another user's and one's own analytic provenance.
In this section, we first describe this prototype, \app, as well as our overall study design(s).

\subsection{\app: System Prototype for the User Study}

With no similar existing system, we developed one wherein users can utilize provenance attributes to perform visual data analysis (e.g., inspect a dataset, create visualizations, and apply transformations) and answer study-specific questions about their provenance. Figure~\ref{fig:teaser} shows the UI with seven views:

\begin{figure*}[ht]
    \centering
    \includegraphics[width=\linewidth]{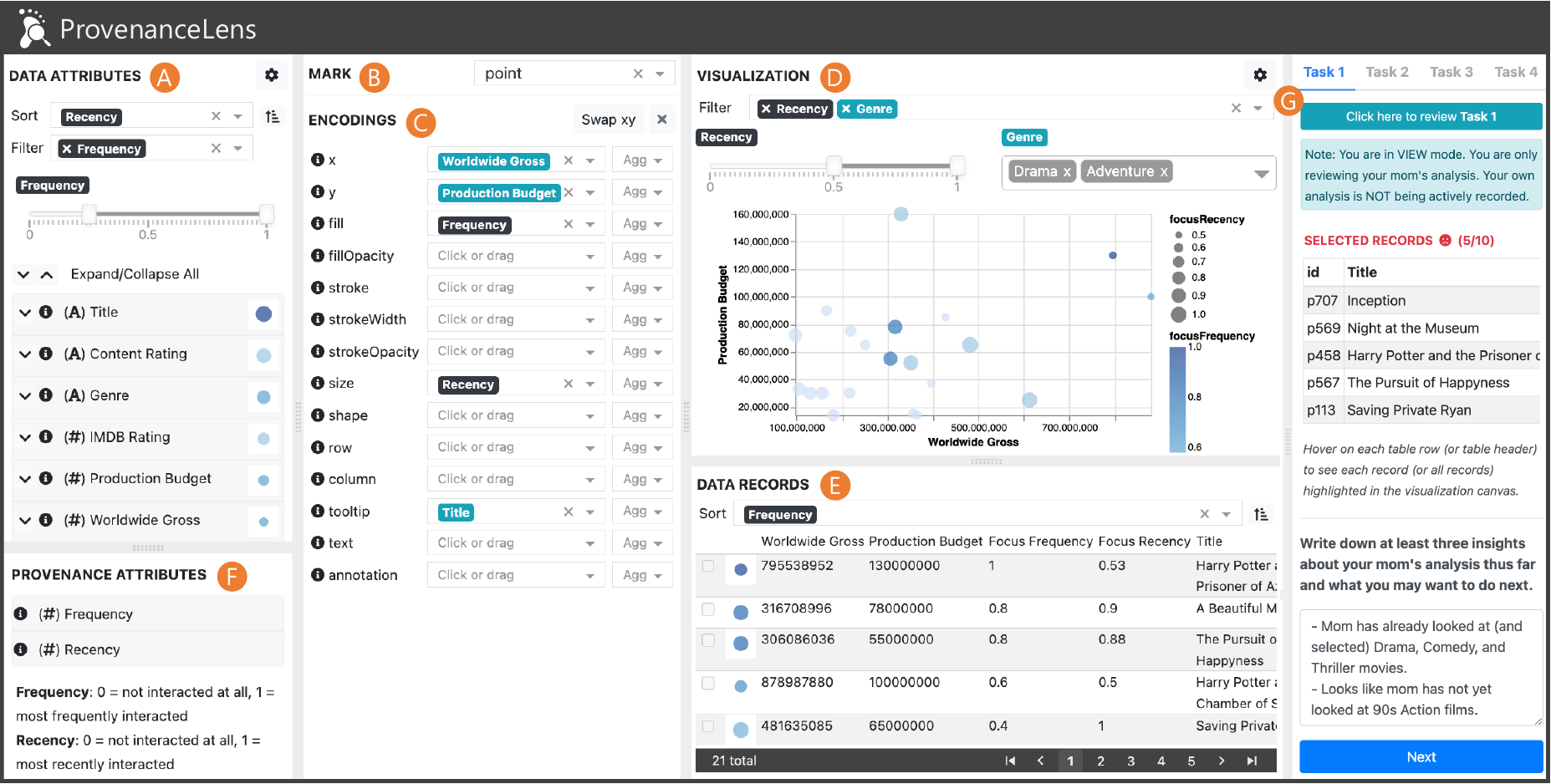}
    \caption{The \app user interface consisting of seven views: \orangecircleglyph{A}the \textbf{Data Attributes} view shows the attributes and enables transformation (e.g., sort, filter); \orangecircleglyph{B}the \textbf{Marks} and \orangecircleglyph{C}\textbf{Encodings} views specify the visualization; \orangecircleglyph{D}the \textbf{Visualization} view renders the specified visualization and supports filtering of data records; \orangecircleglyph{E}the \textbf{Data Records} view supports review and transformation (sort) of the data records shown in the visualization; \orangecircleglyph{F}the \textbf{Provenance Attributes} view lists the \emph{recency} and \emph{frequency} attributes; and \orangecircleglyph{G}the \textbf{Tasks} view shows the task instructions and questions, and tracks the user's progress.}
    \label{fig:teaser}
\end{figure*}

\paragraphHeadingSpace\bpstart{\orangecircleglyph{A}Data Attributes.} In this view, users upload and configure the dataset (\faCog) and see the underlying attributes (or features or columns).
Hovering on the information icon~\faInfoCircle~shows the attribute's definition in a tooltip.
Clicking on the expand icon~\faChevronDown~opens a detailed view with a distribution plot of the attribute's values: an area curve for numerical attributes and a column chart for categorical attributes, both of which show percentage counts corresponding to the attribute quantiles and categories, respectively.
Users can sort and filter the attributes using the \emph{recency} and/or \emph{frequency} provenance attributes.

\paragraphHeadingSpace\bpstart{\orangecircleglyph{B}Mark.} This view includes a dropdown to configure the mark type for the visualization. Users can select one of \emph{point}, \emph{bar}, \emph{line}, \emph{area}, or \emph{text} to begin a visualization specification.

\paragraphHeadingSpace\bpstart{\orangecircleglyph{C}Encodings.} This view shows the visual encoding channels. Users select or drag one or more attributes (data and/or provenance) to one of \emph{x}, \emph{y}, \emph{fill}, \emph{fillOpacity}, \emph{stroke}, \emph{strokeOpacity}, \emph{strokeWidth}, \emph{shape}, \emph{row}, \emph{column}, \emph{tooltip}, and/or \emph{text} to complete a visualization specification. An additional encoding, \emph{annotation}, adds a new annotation displaying the value of the encoded entity next to the selected mark.

\paragraphHeadingSpace\bpstart{\orangecircleglyph{D}Visualization.} This view renders an interactive visualization based on the selected mark type and activated visual encodings in the ``Encodings'' view. It also includes a ``Filter'' drop-zone to filter out data points by attributes (data and/or provenance). A numerical attribute displays a range slider and a categorical attribute displays a multi-select dropdown.

\paragraphHeadingSpace\bpstart{\orangecircleglyph{E}Data Records.} This view shows the data bound in the visualization as a paginated data table.
If the user hovers on a data point in a unit visualization (e.g., a scatterplot), this table filters to only show that data record whereas if the user hovers on an entity in an aggregate visualization (e.g., a bar showing the \emph{mean} value), this table filters to show all data records belonging to the hovered entity (e.g., the bar). 

\paragraphHeadingSpace\bpstart{\orangecircleglyph{F}Provenance Attributes.} This view shows the two provenance attributes: \emph{frequency} and \emph{recency}. Like data attributes, users can select or drag these provenance attributes and drop them to the sort and/or filter drop zones in the ``Attribute View'', the encoding channels in the ``Encodings'' view, the filter drop zone in the ``Visualization'' view, or the sort drop zone in the ``Data Records'' view.

\paragraphHeadingSpace\bpstart{\orangecircleglyph{G}Tasks.} This view has six tabs, one for each task (T1--T6) for study participants to access the current task instructions, track their progress, and answer questions via integrated forms.

\paragraphHeadingSpace\app can be programmatically configured into three modes: (1)~\emph{edit}-only, wherein provenance is tracked and presented back to the user in real-time, (2) \emph{view}-only, wherein existing provenance is imported into \app without real-time tracking, and (3) \emph{hybrid}, wherein provenance is imported into the system, and real-time tracking is also~enabled.

\subsection{Example Usage Scenarios for \app}
\label{section:usagescenarios}
We present two usage scenarios on how \app can enhance visual data analysis for real-time provenance tracking as well as post-analysis review of a user's provenance.

\paragraphHeadingSpace\bpstart{Real-time Provenance Tracking.} Assume Mark works~for a movie production company and must determine what kinds of movies to make next. They upload the dataset of movies in \app (configured in its \emph{edit}-only mode, i.e. real-time provenance tracking) and begin exploring (by specifying different visualizations, applying relevant filters, and hovering on certain movies). 
After taking a short break, they wish to revisit their most recently hovered movie, so they create a visualization with a \emph{point} mark type and map the \emph{recency} provenance attribute to the \emph{x} encoding. They immediately find their desired movie (with the highest recency score) at the horizontal axis' rightmost-end. Had they mapped \emph{recency} to another encoding such as \emph{fill} or \emph{size}, it may have taken them some time to accurately determine the darkest or biggest point.

Upon further inspection of their focus on different movies, Mark noticed they have interacted with some movies multiple times and not considered a lot of other movies (i.e., they exhibited less exploration coverage).
Mark wants to change this analytic behavior, and thus starts visualizing traces of their interactions in real-time.
Because they are colorblind, they avoid the \emph{fill}, \emph{stroke}, \emph{fillOpacity}, and \emph{strokeOpacity} encodings and choose \emph{size} to encode the \emph{recency} provenance attribute. They can now track visited points using their bigger size and continue exploring. 
However, because visited points get bigger, they are getting drawn to the same points even more; thus, they reverse the \emph{size} range to make the visited points smaller instead.
Happy with this configuration, they continue exploring and eventually submit their analysis report to their manager.

\paragraphHeadingSpace\bpstart{Post-Analysis Provenance Review/Audit.} In addition to real-time provenance tracking, \app can also \emph{import} an existing log of a user's analytic provenance to facilitate collaboration (e.g., continuing a colleague's analysis), auditing (e.g., inspecting a colleague's analysis), or post-hoc analysis (e.g., reviewing user study logs~\cite{revisit2023ding,nobre2021revisit}).

Assume Anya is Mark's manager and is reviewing their previous analysis. They express surprise at Mark's recommendation to make a \emph{Drama} movie next.
Wanting to review Mark's analysis, Anya imports Mark's analytic provenance, that was exported from \app, back into \app. They configure \app in its \emph{view}-only mode, i.e.,~no real-time provenance tracking, to review and interact with Mark's analysis.
They make a bar chart with ``Genre'' on \emph{x} and the sum of \emph{frequency} on \emph{y}. This visualization shows Mark's total focus across movie genres. Anya notices that certain bars corresponding to \emph{Action} and \emph{Adventure} movies are really short (i.e., of low \emph{frequency}). They ask Mark to review those genres before making a final recommendation.
In this way, Anya was not only able to review Mark's report but also their analysis process, thereby supporting enhanced decision-making.

\subsection{Pilot Studies and Evaluation Considerations}
\label{section:pilotstudies}
Before finalizing our study design, we explored two alternate designs and also conducted subsequent pilot studies.

\paragraphHeadingSpace\bpstart{Pilot Study 1: Decision-Making.} We recruited four Ph.D. students (three years into the program) as our pilot users.
We tasked users to explore a dataset of movies and select (1)~ten movies (records) satisfying certain criteria, (2)~four movie characteristics (attributes) that were important to their analysis, and (3)~answer a series of questions about their analysis.

We observed users did not utilize the provenance attributes to track their analysis process and only used them during the subsequent question-answering. We also noticed that answering questions immediately after analysis may not be hard for certain users because the analysis may be quite fresh in their memory. In addition, the idea to make users select important attributes was straightforward and seemed redundant because the task criteria already hinted what attributes to consider.

\paragraphHeadingSpace\bpstart{Pilot Study 2: Analysis Review and Decision-Making.} We recruited two other Ph.D. students (four years into their program) as our pilot users.
In this revised study design, we introduced an initial \emph{review} task to first make the user explore another user's analytic provenance, write three insights, and then answer questions about the prior analysis process. The original task to select movies followed this task. Our goal was to make the user actively utilize the provenance attributes during their own analysis, and we believed making them answer questions about another user's analysis first would remove the fresh-in-memory aspect from the equation.
We also discarded the selection of four movie characteristics.
Overall, we noticed this study design had a desired positive effect and decided to use it for our eventual user study, described next.

\subsection{Exploratory User Study Design}
\bpstart{Participants:} We recruited 16 participants from a public university in the U.S. who were pursuing a bachelors (1), masters (11), or doctoral (4) degree in computing or related fields (15) and economics (1). These participants were either enrolled in or alumni of at least one data visualization class and self-reported their visualization literacy to be at least 3 on a scale from 1 (novice) to 5 (expert). Demographically, they were in the \emph{18-24}~(7) or \emph{25-34}~(9) age groups (in years) and of \emph{female}~(5), \emph{male}~(10), or \emph{preferred not to say}~(1) genders.

\paragraphHeadingSpace\bpstart{Study Session:} 
Each study session lasted between 75 and 90 minutes. We compensated each participant with a \$15 gift card for their time. We conducted the study remotely over Zoom; the experimenter provided participants access to the study environment by sharing their (the experimenter's) computer screen and granting input control to the participant. After providing consent, participants saw a five-minute video tutorial that demonstrated the features of \app. Participants then performed a practice task on a \emph{dataset of cars} to get acquainted with the UI before starting the actual task.

The actual task was on a dataset of \emph{movies} and lasted about 60 minutes. Participants were not required to think aloud during the task to simulate a realistic work setting (although some participants felt comfortable doing so). During the task, participants' interactions with the system were logged. The study ended with participants completing a feedback questionnaire and a background questionnaire. Each study session was screen- and audio-recorded for subsequent analysis.

\paragraphHeadingSpace\bpstart{Task and Dataset:} 
We designed the following visual data exploration and decision-making task about a movies dataset:

\begin{highlight}
\small Imagine your family is planning a month-long vacation to Europe. Going with you are your siblings, parents, grandparents, your uncle and aunt, and their boy (your cousin). Your mom began selecting some movies to pick and watch from for the occasional movie nights. \textbf{Her target is to select ten movies to carry to the vacation}.

\paragraphHeadingSpace\noindent Because she wanted to ensure a delightful and well-rounded movie night experience for your entire family, \textbf{she took the below suggestions and preferences from some of your family members into consideration}.
\paragraphHeadingSpace
\begin{enumerate}[nosep, leftmargin=*]
    \item Dad: \emph{``I like thrillers and comedies.''}
    \item Uncle: \emph{``Let's watch a heartfelt drama.''}
    \item Cousin: \emph{``I want to re-watch a Harry Potter movie.''}
    \item Grandpa: \emph{``I would love to watch a 90s' action film.''}
    \item Aunt: \emph{``I want to watch a hidden gem: a highly-rated movie that didn't do well commercially.''}
\end{enumerate}

\paragraphHeadingSpace\noindent\textbf{After selecting five movies}, your mother got pulled to do another task, and assigned you to complete it.

\paragraphHeadingSpace\noindent\textbf{Thus, you will complete the following six tasks (\textbf{T1-T6}).}

\paragraphHeadingSpace\emph{------ Working with your mom's analytic provenance ------}
\begin{enumerate}[nosep, leftmargin=0.5cm]
    \item[T1] \textbf{Review:} Review the five movies your mom selected and many others she explored; write three insights.
    \item[T2] \textbf{Recall:} Answer five objective questions about mom's analysis.
    \item[T3] \textbf{Visualize:} Create visualizations to answer subjective questions about mom's analysis.
\end{enumerate}

\paragraphHeadingSpace\emph{------ Working with your own analytic provenance ------}
\begin{enumerate}[nosep, leftmargin=0.5cm]
    \item[T4] \textbf{Analyze:} Select the remaining five movies.
    \item[T5] \textbf{Recall:} Answer five objective questions about your analysis.
    \item[T6] \textbf{Visualize:} Create visualizations to answer subjective questions about your analysis.
\end{enumerate}

\paragraphHeadingSpace\noindent To help you review and track the movies (records) and their characteristics (attributes) one has \emph{interacted} with how many times and when, our system provides two special provenance attributes: \textbf{Frequency} and \textbf{Recency}. You can use these in the same way you would use the dataset attributes, i.e., create visualizations by mapping them to visual encodings, filter by them, or sort by them.

\paragraphHeadingSpace\noindent Note that ``\emph{interacted}'' refers to a user's interactions in the interface such as hovering on a record to get additional details, 
mapping dataset attributes and/or provenance attributes to visual encodings (e.g., `Genre' to \emph{x}), applying a filter (e.g., `Title'=`Titanic') or sort.
\end{highlight}

\definecolor{c2.38}{RGB}{207, 243, 222}
\definecolor{c2.69}{RGB}{196, 241, 215}
\definecolor{c2.56}{RGB}{201, 242, 218}
\definecolor{c1.94}{RGB}{222, 247, 233}
\definecolor{c4.19}{RGB}{144, 228, 180}
\definecolor{c1.63}{RGB}{233, 250, 240}
\definecolor{c3.56}{RGB}{166, 233, 194}
\definecolor{c1.88}{RGB}{224, 248, 234}

\definecolor{c100}{RGB}{46, 204, 113}
\definecolor{c93.75}{RGB}{59, 207, 122}

\definecolor{c6.81}{RGB}{52, 206, 118}
\definecolor{c6.75}{RGB}{54, 207, 119}
\definecolor{c6.69}{RGB}{56, 207, 120}
\definecolor{c6.06}{RGB}{79, 213, 135}
\definecolor{c6.88}{RGB}{51, 206, 117}
\definecolor{c6.25}{RGB}{69, 211, 128}
\definecolor{c6.38}{RGB}{65, 210, 125}
\definecolor{c6.63}{RGB}{58, 208, 121}
\definecolor{c6.44}{RGB}{63, 209, 124}

\renewcommand{\arraystretch}{1}

\begin{table*}[t]
    \footnotesize
    \setlength{\tabcolsep}{2pt}
    \begin{tabular}{lllrrrrr}
        \toprule
        \textbf{Task} & \textbf{Que} & \textbf{Description} & \textbf{$\mu$Int.} & \textbf{$\mu$Time} & \textbf{$\mu$Acc.} & \textbf{$\mu$Conf.} & \textbf{$\mu$Sur.} \\
         &  &  &  & (mins) & (\%age) & (/7) & (/7) \\
        \midrule
        T1 (Review) & Q1 & Review YOUR MOM's analysis and write three insights. & 53.43 & 7.32 & - & - & - \\
        \midrule
        T2 (Recall) & Q1 & Select the movie characteristic(s) that YOUR MOM interacted with the MOST. & 12.69 & 1.35 & \cellcolor{c100}{100} & \cellcolor{c6.81}{6.81} & - \\
         & Q2 & Which was the last (i.e. most recent) movie YOUR MOM interacted with? & 15.88 & 2.32 & \cellcolor{c93.75}{93.75} & \cellcolor{c6.81}{6.81} & - \\
         & Q3 & Did YOUR MOM ever interact with the movie `Titanic'? & 12.69 & 1.89 & \cellcolor{c100}{100} & \cellcolor{c6.75}{6.75} & - \\
         & Q4 & Did YOUR MOM interact with at least one movie from all `Content Rating's? & 13.38 & 1.94 & \cellcolor{c100}{100} & \cellcolor{c6.69}{6.69} & - \\
         & Q5 & Did YOUR MOM interact with all attributes at least once? & 17.88 & 1.49 & \cellcolor{c100}{100} & \cellcolor{c6.75}{6.75} & - \\
        \midrule
        T3 (Visualize) & Q1 & What was the distribution of YOUR MOM's focus across different movie `Genre's? & 12.13 & 2.08 & \cellcolor{c100}{100} & \cellcolor{c6.75}{6.75} & - \\
         & Q2 & How did YOUR MOM's focus on `Drama' Movies evolve over time? & 35.50 & 4.22 & \cellcolor{c100}{100} & \cellcolor{c6.06}{6.06} & - \\
         & Q3 & Which were YOUR MOM's most FREQUENTLY interacted movies? Try to show five. & 24.69 & 2.69 & \cellcolor{c100}{100} & \cellcolor{c6.63}{6.63} & - \\
        \midrule
        T4 (Analyze) & Q1 & Select the five remaining movies. & 89.75 & 7.24 & - & - & - \\
        \midrule
        T5 (Recall) & Q1 & Select three movie characteristic(s) YOU most RECENTLY interact with. & 13.75 & 1.90 & \cellcolor{c100}{100} & \cellcolor{c6.88}{6.88} & \cellcolor{c1.88}{1.88} \\
         & Q2 & Which was the first (i.e. earliest) movie YOU interacted with? & 22.13 & 3.01 & \cellcolor{c93.75}{93.75} & \cellcolor{c6.25}{6.25} & \cellcolor{c3.56}{3.56} \\
         & Q3 & Did YOU interact with the movie `Pearl Harbor'? & 8.63 & 1.44 & \cellcolor{c100}{100} & \cellcolor{c6.81}{6.81} & \cellcolor{c1.63}{1.63} \\
         & Q4 & Which `Content Rating' category did YOU interact with the LEAST or NONE AT ALL? & 11.25 & 1.98 & \cellcolor{c100}{100} & \cellcolor{c6.69}{6.69} & \cellcolor{c4.19}{4.19} \\
         & Q5 & Did YOU interact with at least half of the movie characteristics available? & 12.38 & 1.22 & \cellcolor{c100}{100} & \cellcolor{c6.88}{6.88} & \cellcolor{c1.94}{1.94} \\
        \midrule
        T6 (Visualize) & Q1 & How similar were YOUR interaction patterns for `Comedy' and `Thriller' movies? & 17.81 & 2.67 & \cellcolor{c100}{100} & \cellcolor{c6.38}{6.38} & \cellcolor{c2.56}{2.56} \\
         & Q2 & Which were YOUR most RECENTLY interacted movies? Try to show THREE movies. & 19.00 & 2.42 & \cellcolor{c100}{100} & \cellcolor{c6.63}{6.63} & \cellcolor{c2.69}{2.69} \\
         & Q3 & Given an opportunity, which (kinds of) movies would YOU like to go back and interact with? & 18.88 & 2.74 & \cellcolor{c100}{100} & \cellcolor{c6.44}{6.44} & \cellcolor{c2.38}{2.38} \\
        \bottomrule
        \\
    \end{tabular}
    \caption{Tasks and summary performance statistics for sixteen participants: \textbf{Task} and \textbf{Que}stion index, task \textbf{Description}, average number of interactions (\textbf{$\mu$Int.}) and time spent in minutes (\textbf{$\mu$Time}), and wherever applicable,  average accuracy (\textbf{$\mu$Acc.}), average confidence (\textbf{$\mu$Conf.}), and average surprise (\textbf{$\mu$Sur.}) on a scale from one (low) to seven (high). 
    T1 and T4 were exploratory in nature, hence we did not compute accuracies or ask participants to self-report confidence and surprise scores.
    Similarly, T2 and T3 were focused on answering questions about mom's analysis, hence we did not ask participants to self-report their surprise scores.
    Overall, participants performed exceedingly well on all the tasks, achieving high accuracies with high confidence with some moments of surprise.}
    \label{tab:task-questions}
\end{table*}

\section{Results}
\label{section:results}
In this section, we present general and task-specific findings from the user study along with participant performance on the six tasks (T1-T6) and discuss them in the context of the qualitative feedback from our participants (\user{1,...,16}).
Detailed participant responses and performance assessments for tasks T1-T6, post-study feedback questionnaire and responses, and a demo video of the UI are available in supplemental material.

\subsection{General Feedback}
Participants scored \app 80.94 out of 100 on the system usability scale (SUS~\cite{brooke1996sus}), finding it very useful.
\user{15} acknowledged it is impossible to remember everything during analysis, calling for \emph{``Tools like Tableau [to] really overlay communities' or experts' interaction stats such as frequency and recency in their system. It is one of those features that only adds value and is not necessarily a hindrance.''}
\user{14} was impressed that the two provenance attributes could help answer a variety of questions, enabling users to revisit, review, and maybe even recreate someone's history of interactions while doing analysis.
\user{11} liked being able to detect subconscious interaction patterns during analysis.
\user{16} suggested a potential use case, \emph{``If you are a manager and if you need to review the decision process of your employees, then it is very useful.''}

\subsection{Task Performance}
\label{sec:accuracy}
Table~\ref{tab:task-questions} shows the task-specific breakdown of participants' accuracy and self-reported confidence and surprise, along with the number of interactions and time spent during the study.

\paragraphHeadingSpace\bpstart{Accuracy.} 
We computed accuracy by comparing participant's responses to each question with corresponding answers computed from their actual interaction behavior (ground truth). To ensure consistency and reliability, we automated this process using a standardized Python script, as the questions were fixed across all participants.
Participants performed exceedingly well overall, achieving high accuracies during the \textbf{recall} tasks (T2, T5) and the \textbf{visualize} tasks (T3, T6). 
The mean accuracies were as follows: T2 ($\mu=98.75\%$), T3 ($\mu=100\%$), T5 ($\mu=98.75\%$), T6 ($\mu=100\%$).
We did not compute accuracy for the \textbf{review}~(T1) and \textbf{analyze}~(T4) tasks as they were exploratory in nature.

Only two participants~(\user{13, 14}) incorrectly answered one question each.
For instance, for T2.Q2 (\emph{``Which was the last (i.e. most recent) movie YOUR MOM interacted with?''}), \user{14} selected the `point' mark type and assigned ``Genre'' to \emph{x}, \textit{recency} on \emph{size}, and ``Title'' on \emph{tooltip}. Then, they applied a \textit{recency} filter of [0.63, 1]. This configuration resulted in four marks stacked on top of each other (unknown to the participant who thought there was only one). The user hovered on it, read \emph{``The Curious Case of Benjamin Button''} in the tooltip, and selected it as the answer. However, this answer was incorrect because \emph{``Saving Private Ryan''}, under this movie, had the higher recency score of 1.0. This error may have been avoided if instead of (or along with) \emph{size}, the user assigned \textit{recency} to \emph{x} or \emph{y}, which would have spaced out the points.

\paragraphHeadingSpace\bpstart{Confidence \& Success.} On a scale from 1 (low) to 7 (high), participants self-reported very high \emph{confidence} in answering questions during tasks T2 ($\mu$=6.76, M=7), T3 ($\mu$=6.48, M=7), T5 ($\mu$=6.70, M=7), and T6 ($\mu$=6.48, M=7). 
In addition, while answering questions based on their own analysis, participants self-reported varying \emph{surprise} during tasks T5 ($\mu$=2.64, M=2) and T6 ($\mu$=2.54, M=2).
Note that T1 and T4 were exploratory in nature, hence we did not ask participants to self-report confidence and surprise scores. 
Similarly, T2 and T3 were focused on answering questions about mom's analysis, hence we did not ask participants to self-report their surprise scores.

\paragraphHeadingSpace\bpstart{Fidelity.} On a scale from 1 (low) to 5 (high), participants self-reported that overall, all six tasks (T1-T6) caused low \emph{physical demand} ($\mu$=1.44, M=1), \emph{temporal demand} ($\mu$=2.44, M=2), and \emph{frustration} ($\mu$=1.5, M=1), average \emph{mental demand} ($\mu$=3.31, M=3) and \emph{effort} ($\mu$=2.94, M=3), but resulted in high \emph{performance} ($\mu$=4.38, M=4). 
On average, participants spent around 56 minutes completing the study and performed 409 interactions.
We did not seek feedback after each task to avoid demotivating the user's analysis with frequent interruptions.

\paragraphHeadingSpace\bpstart{Summary.} The high \emph{accuracy}, \emph{confidence}, and \emph{performance} along with average \emph{effort} and \emph{mental demand} suggest that participants were able to effectively use provenance attributes to answer the questions.
The high variance in surprise scores suggests that participants were sometimes able to recall their analysis from memory (less surprise) or had an incorrect mental model or recall of their interactions (more surprise).

\subsection{Task-specific Feedback}

\bpstart{Reviewing Mom's Previous Analysis (Review: Task T1).}
Due to the nature of this task, all participants leveraged the provenance attributes and found them to be useful.
Participants felt that the provenance attributes could provide \emph{``a record of [mom's] analysis''} (\user{12}), which made it easier to \emph{``track her way of thinking through a quantitative behavior analysis''} (\user{3}). 
For \user{6}, the provenance attributes were only somewhat useful because they were able to understand mom's analysis simply by inspecting her final selections. 
On the other hand, \user{4} argued that the provenance attributes \emph{``gave context for [mom's movie selections],''} which they would have to otherwise guess (\user{12}).

\paragraphHeadingSpace\bpstart{Answering Questions about Both Mom's and One's Own Analysis (Recall, Visualize: Tasks T2, T3, T5, T6).}
Like T1, due to the nature of the questions, all participants found the provenance attributes to be useful. 
\user{1} noted, \emph{``It was very useful to use provenance attributes as filters and encodings. I would not have been able to answer questions effectively without the visualization ability.''}
\user{14} noted, \emph{``While looking at and analyzing [mom's] focus, it was hard to keep all the insights in mind, so having it visualized helped understand and hence remember it better.''}
\user{9} found the provenance attributes to be more helpful to answer questions about visualizing focus (T3, T6) than searching for insights (T2, T5), noting that \emph{``I was able to concentrate on specific facets of [mom's] focus at a time [via the visualizations], which felt more organized.''}

\paragraphHeadingSpace\bpstart{Selecting Remaining Movies (Analyze: Task T4).}
Ten participants (\user{1,2,3,5,6,7,10,11,12,16}) either did not use the provenance attributes or found them to be less useful. Among these, \user{3} was more focused on selecting movies that satisfied the family's constraints and hence was more inclined to check the actual attributes rather than the \emph{frequency} and \emph{recency}.
\user{6, 12} speculated that the provenance attributes would be more useful if the study lasted longer or was spread across multiple~sessions.

On the other hand, six participants (\user{4,8,9,13,14,15}) found utility in tracking and reviewing their provenance in real-time.
\user{13} actively used the provenance attributes to 
keep track of their own analysis and not become stuck while making a choice.
\user{4} realized that after satisfying all family members' movie constraints, they were free to give their own recommendations. They saw they had not interacted with ``Production Budget'' and ``Worldwide Gross'' too much, so they created a scatterplot with these attributes, mapped \emph{frequency} and \emph{recency} on \emph{color} and \emph{size}, and utilized the visual scents to analyze unexplored movies and accordingly determine their final movie selections.

\subsection{Reasons for Using the Provenance Attributes}
\label{subsection:reasonsforusingprovenanceattributes}

\bpstart{Answer questions when unable to recall analysis behavior.} 
Due to the study design, answering questions was the most common reason to use the provenance attributes.
Reviewing analytic behavior as a quantized set of variables (\user{3}) and using them in visual encodings (\user{1, 14}) and data transformations (\user{1}) made it easy to confidently answer questions (\user{3}).

\paragraphHeadingSpace\bpstart{Verify and build confidence while answering questions.}
Sometimes participants recalled answers from memory, but using the provenance attributes helped them verify and double-check their choices (\user{14, 16}) and gain confidence (\user{12,14}).

\paragraphHeadingSpace\bpstart{Save time during analysis.} 
Some participants found the \emph{recency} and \emph{frequency}-based data transformations and aggregations particularly useful to generate quick responses to the study questions (\user{11}), directly saving them a lot of time (\user{10}).

\paragraphHeadingSpace\bpstart{Increase awareness, uncover new insights, and improve exploration coverage.}
When selecting the remaining movies, the provenance attributes helped participants optimize their analysis by increasing awareness (\user{4}), helping~them avoid revisiting the same points (\user{9,15}), and facilitating quick decisions~(\user{15}).
Some participants uncovered new insights~(\user{2, 3}) or unexplored data~(\user{3}), and identified behavioral trends (\user{3}).

\paragraphHeadingSpace\bpstart{Fun and Feel Good.} For some participants, the provenance attributes were easy and fun to use (\user{13}). The real-time changes to the colors or sizes of points made them feel good and feel like they were making progress in the task~(\user{13}).

\subsection{Reasons for More or Less Surprise}
While answering questions, participants were surprised because (1) the system's answer did not match their recollection of their own analysis, (2) the system's answer did not match their original analytic intention, (3) the system inadequately captured the participant's focus (e.g., interactions with an aggregate visualization gave equal focus to constituent datapoints, which was not agreeable to the user), or (4) the system logged interactions that were perhaps accidental (our threshold to discard mouseovers under \(\sim \)250ms may be low for use as a proxy to compute focus). 
For example, \user{2} noted, \emph{``I found [an interaction] to be a Harry Potter movie, which makes sense, but I had forgotten about having interacting with it first.''}

On the flip side, participants were either less surprised or not surprised at all because (1) they actively used the recency and frequency attributes during analysis and thus knew what they were doing and (2) they were able to simply remember their behaviors.
\user{5} noted, \emph{``I already knew what I had focused on, but far ahead in the future, these provenance attributes might be helpful to provide reasoning for my choices when I might not explicitly remember my thought process.''}

\begin{figure*}[t]
    \centering
    \includegraphics[width=\linewidth]{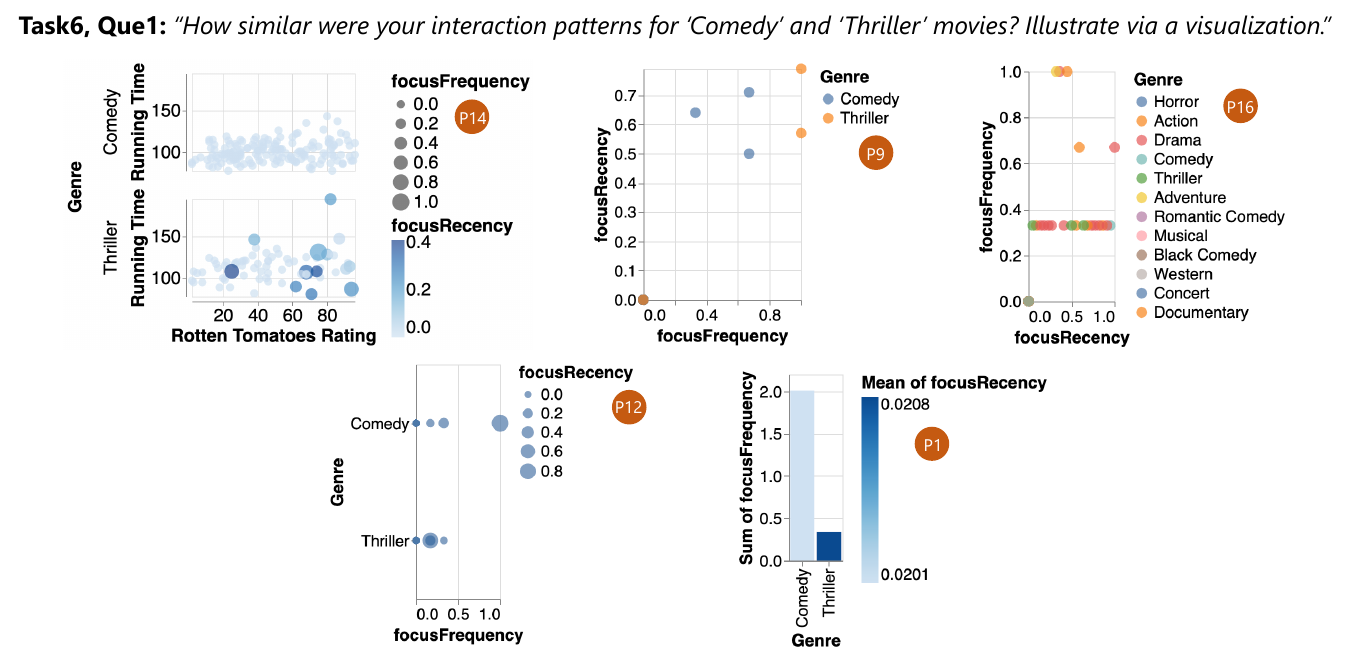}
    \caption{Five participants' different strategies to answer the same question, T6.Q1, \emph{``How similar were your interaction patterns for `Comedy' and `Thriller' movies? Illustrate via a visualization.''} \user{14} created a scatterplot with ``Rotten Tomatoes Rating'' along \emph{x}, ``Running Time'' along \emph{y}, faceted by ``Genre'', colored by \emph{recency}, and sized by \emph{frequency}; \user{9} and \user{16} created a scatterplot with \emph{frequency} and \emph{recency} along \emph{x} or \emph{y}, and colored by ``Genre''; \user{12} created a scatterplot with \emph{frequency} along \emph{x}, ``Genre'' along \emph{y}, and sized by \emph{recency}; \user{1} created an aggregate bar chart with ``Genre'' along \emph{x}, total \emph{frequency} along \emph{y}, and colored by average \emph{recency}. Except \user{16}, each of the above participants also applied a filter to only show movies that belong to the ``Comedy'' or ``Thriller'' genre.}
    \label{fig:quiz-answers-task6}
\end{figure*}

\subsection{Participant Strategies to Answer Questions}

To answer questions, participants often utilized provenance attributes as a combination of encodings, transformations (sort and filter), and subsequent interactions (e.g., tooltip on hover). We observed 71 different strategies. For example, Figure~\ref{fig:quiz-answers-task6} shows five participants' different strategies to answer T6.Q1: \emph{``How similar were your interaction patterns for `Comedy' and `Thriller' movies?''} While \user{1} created an aggregate bar chart showing the two ``Genre'' categories on \emph{x}, \emph{frequency} on~\emph{y}, and average \emph{recency} as \emph{color}, \user{14} used a scatterplot encoding \emph{recency} (color) and \emph{frequency}~(size), faceted by ``Genre''.

\begin{figure*}[t]
    \centering
    \includegraphics[width=\linewidth]{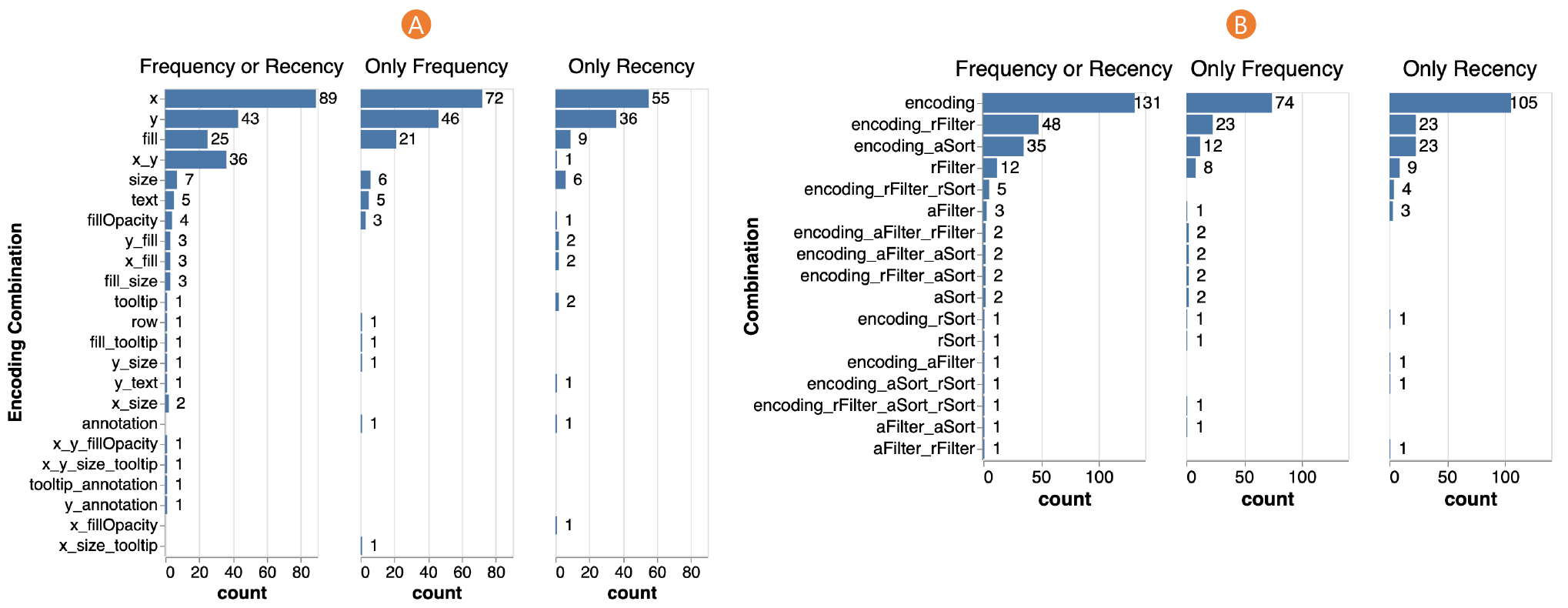}
    \caption{Co-occurrence statistics for how users map provenance attributes (\emph{only frequency}, \emph{only recency}, or either) to visual encoding combinations (A), as well as general preferences for visual encodings compared to filtering, and sorting (B). Note that these statistics correspond only to the \textbf{recall (T2, T5)} and \textbf{visualize (T3, T6)} tasks; we exclude the \textbf{review (T1)} and \textbf{analyze (T4)} tasks as they were more open-ended in nature.}
    \label{fig:cooccurrence_stats}
\end{figure*}

\paragraphHeadingSpace\bpstart{Co-occurrence analysis.} Figure~\ref{fig:cooccurrence_stats} shows the co-occurrence of provenance attributes (\emph{only frequency}, \emph{only recency}, or either) as \orangecircleglyph{A}\kern0px combinations of visual encodings compared to \orangecircleglyph{B}\kern0px data transformations.
In terms of visual encodings, \emph{x} and \emph{y} independently were used most often followed by \emph{fill} (color). 
Interestingly, \emph{frequency} and/or \emph{recency} were simultaneously mapped to both \emph{x} and \emph{y} a solid 36 times.
Between visual encodings and data transformations, standalone visual encodings were used most often (131 times) followed by their combination with record filter (\emph{encoding\_rFilter}, 48 times) and attribute sort (\emph{encoding\_aSort}, 35 times).

\subsection{Preferences for Utilizing the Provenance Attributes}
\begin{figure*}[t]
    \centering
    \includegraphics[width=\linewidth]{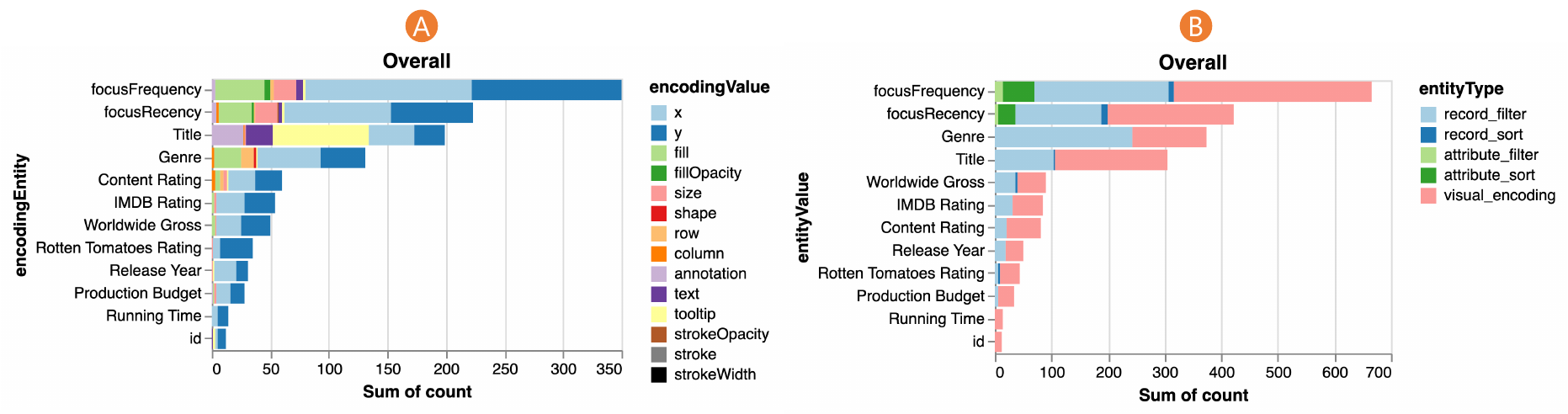}
    \caption{User preference when mapping attributes to (A) different visual encodings and (B) using visual encodings in general compared to data transformations.}
    \label{fig:popularity-stats}
\end{figure*}

Figure~\ref{fig:popularity-stats} shows how users mapped data and provenance attributes to different visual encodings~\orangecircleglyph{A}\kern-3px, and their general preferences for encodings compared to transformations~\orangecircleglyph{B}\kern-3px. 

\paragraphHeadingSpace\bpstart{Visual Encodings.} 
For provenance attributes, \emph{x} was the most preferred encoding followed by \emph{y}, \emph{fill}, \emph{size}, \emph{text}, \emph{fillOpacity}, \emph{row}, \emph{annotation}, and \emph{column}. The encodings \emph{shape}, \emph{stroke}, \emph{strokeOpacity}, and \emph{strokeWidth} were either rarely used or not used at all.
Similarly, for data attributes, \emph{x} or \emph{y} were the most preferred encodings. The nominal ``Genre'' and ``Content Rating'' attributes were also occasionally mapped to \emph{fill} and \emph{row} encodings. ``Title'' was predominantly mapped to the \emph{tooltip} visual encoding, followed by \emph{x}, \emph{annotation}, \emph{y}, and \emph{text}.

\paragraphHeadingSpace\bpstart{Visual Encodings or Data Transformations.} 
For provenance as well as data attributes, visual encodings were most preferred followed by filtering records and sorting attributes.

\subsection{Recency or Frequency? What was more Useful?}

Three participants (\user{1, 5, 14, 16}) highlighted that frequency was more useful than recency. \user{14} mentioned that they did not rely much on recency for selecting movies but found frequency helpful in identifying movies they had visited multiple times. 
\user{11} also noted that
frequency aided in recognizing options that were still being considered.
\user{15} suggested that while frequency was more useful for the short duration of the study, recency might be more valuable in the long term, especially for tasks like auditing or reviewing analysis after days or weeks.

\section{Discussion and Takeaways}
\label{section:discussion}
\subsection{Making Provenance a Core Aspect of Analysis}
Our study revealed that users can use provenance attributes to answer questions with high accuracy and confidence, while also sometimes being surprised. 
In addition, there were several instances of users requesting new capabilities not currently supported by \app. 
For example, \user{6} suggested an interesting feature to toggle between their own provenance and their mom's during analysis. 
\user{8} requested the ability to ``undo'' an accidental interaction, hinting towards an ability to directly manipulate/correct their provenance.
These feature requests solidify that provenance still has unrealized utility, and should thus be considered as a core element during analysis, thereby calling for visual data analysis tools to inherently support it.

\subsection{Fostering Provenance-driven (Not Data-driven) Analysis.}
\label{subsection:foster-prov-analysis}
Many participants visualized provenance by mapping it to visual encodings (e.g., color) to keep track of their ongoing analytic progress (e.g., to avoid revisiting the same points).
Some participants also actively sorted and filtered the data attributes and records by provenance attributes to either reduce their search space or look-up what has been previously considered. 
These are examples of provenance-driven (not data-driven) analysis wherein provenance information is used to steer the analysis process.
Similar to how data-driven analysis focuses on using data to guide decision-making and insight generation, provenance-driven analysis can surface the lineage and context of data and interactions to determine next steps~and ensure the accuracy, reliability, and interpretability of the analysis.
We thus call for tools to support both these analysis paradigms.

\subsection{Integrating Provenance-Tracking \& Visual Data Analysis}
\label{subsection:integrate-prov-visual-data-analysis}
During our study, participants often used provenance attributes together with data attributes. For example, participants mapped provenance attributes to \emph{x}, \emph{y}, \emph{fill}, and \emph{size}, which are encodings commonly used for data attributes.
Furthermore, there were cases where a participant replaced a data attribute (that was mapped to a visual encoding), with a provenance attribute (and vice-versa).
In other cases, participants had a preferred way of interacting with analytic provenance (e.g., always mapping provenance to \emph{fill}) and they consistently used the same strategies during analysis or when answering questions.
These behaviors suggest that tools should offer both data and provenance attributes for more flexible~workflows.

\subsection{Affording Flexibility in Mapping Provenance to Encodings}
Our study revealed that users often visualized provenance on visual encodings such as \emph{x}, \emph{y}, and \emph{tooltip}, something not often seen in existing tools.
Furthermore, while performing tasks T2 (Recall) and T4 (Analyze), some participants used multiple visual encodings for the same question in succession, to verify the results. 
For example, \user{4} first mapped \emph{frequency} to \emph{fill}, but found it hard to accurately differentiate between different shades and hues (which can be hard for colorblind users). 
To verify their takeaways, they mapped \emph{frequency} to \emph{size} instead, but were worried about occlusion, depending on the x/y point positions. 
Consequently, they applied a double-encoding by also mapping \emph{frequency} to \emph{x}.
This flexibility shows that if an encoding is unavailable and another is inferior, then a third encoding can still be effective, underscoring our core goal to afford flexibility in mapping provenance during analysis.

\subsection{Comparing Provenance Encodings and Transformations}
During our study, participants regularly mapped provenance to visual encodings or applied data transformations. However, depending on the question type, one technique can be more efficient than the other. Consider T6.Q2 wherein participants were tasked to show \textbf{three movies} that they most recently interacted with. Those who applied a \emph{recency} filter often struggled with this task due to ties in the number of times a movie was interacted with. For example, \user{1} asked, \emph{``How do I get exactly three?''} Filtering for top-N movies (records) or movie characteristics (attributes) is less trivial as it is impossible to guess what range will produce the exact number of items. 
Mapping provenance to positional encodings (e.g.,~\emph{x}) or filtering based on ordinal \emph{provenance rank} instead of a quantitative \emph{provenance score} can make it easier to spot ties.

\subsection{Supporting Collaboration during Visual Data Analysis}
In tasks T1--T3 of our user study, participants had to utilize provenance attributes to review and answer questions about another user's (mom's) analysis. In doing so, we indirectly studied how provenance attributes can facilitate asynchronous collaboration.
By analyzing another user's provenance in terms of what they looked at, when, and for how long, the user can not only verify or find flaws in prior analyses but also formulate a starting point or become unstuck~\cite{narechania2023datapilot}. 
However, overreliance on another user can hamper creativity and result in ``herd behavior''~\cite{narechania2023datapilot}.
Balancing these two behaviors can result in efficient collaborative analysis and decision-making.

\subsection{Enhancing Modeling of Provenance Attributes}
\label{subsection:implicationsofmodelingprovenanceattributes}
Our approach of modeling provenance attributes has some implications and limitations.
First, mapping \emph{recency} and \emph{frequency} to visual encodings, such as darker and larger points for more recent or frequent data, can foster confirmation bias~\cite{nickerson1998confirmation} by leading users to unknowingly and disproportionately prioritize recently or frequently interacted data points. Such behavior is an unfortunate consequence of the recency effect~\cite{atkinson1968human} and the frequency effect~\cite{tversky1973availability}. As a result, users may potentially overlook less accessed but relevant information. To mitigate this bias, a reversed provenance scale that emphasizes older and less frequent data (as darker and larger) can provide a more balanced view in decision-making processes.

Next, our approach to normalize recency and frequency values to a uniform ranked scale from zero to one, instead of displaying absolute values like raw timestamps and interaction counts, can simplify comparisons but sacrifice explainability. 
For instance, without timestamps, users may not understand the exact sequence or timing of interactions, i.e., whether frequent interactions occurred closely together or were spread out over time.
Additionally, users may overlook the original context (i.e., the data attributes and records themselves) or the rationale of the previous user whose provenance they are reviewing.
Such lack of clarity can lead to incomplete or biased interpretations in decision-making.

Lastly, provenance attributes can be modeled in multiple ways. 
For \emph{frequency}, the default \emph{relative} strategy divides the ``total interaction units'' for an attribute by the maximum value among all attributes; the \emph{absolute} strategy divides each value by the sum of all values; and the \emph{binary} strategy treats zero interactions=0 and at least one interaction=1.
For recency, the default \emph{relative} strategy determines the value based on the sequence (or rank) of interactions; the \emph{absolute} strategy determines the value based on the actual time duration between interactions; and the \emph{binary} strategy assigns a value=1 for the most recent interaction and a value=0 for all other interactions.
Our conceptualization of provenance attributes seamlessly supports all of these models for subsequent visualization.

\subsection{Facilitating Co-Adaptive Guidance during Analysis}
In this study, we studied how users can flexibly map provenance attributes to different encodings and data transformations. We also believe our system can be extended to enable co-adaptive guidance~\cite{sperrle2021co}, wherein the system (not user) can recommend provenance mappings as orienting guidance~\cite{ceneda2016characterizing} to draw the user's attention to specific analytic behaviors such as exploration biases~\cite{narechania2022lumos}.
In response, the user can accept, reject, or override these recommendations, so that the system can learn the user's preferences to inform future analytic discourse. 
In this way, users and systems can work together to provide tailored help, thereby improving the analysis processes.

\subsection{Reflecting on Provenance-focused User-Study Designs}
In our two pilot studies, participants generally ignored the provenance attributes during analysis, only using them later for subsequent question-answering. This behavior was likely due to the task's nature or participants' ability to simply recall their analysis.
In response, we added the ``review'' (your mom's) task and the ``answer questions'' (about your mom's analysis) task to force participants to interact with the provenance attributes that are not modeled on their own analysis, which helped a bit.
Still, during the actual study, \user{12} noted, \emph{``My memory still works pretty well in the short time frame,''} and \user{6} speculated, \emph{``Maybe if it were a longer session or spread across multiple sessions, then [provenance attributes] would be more useful.''}
One key insight in designing such short-duration user studies on provenance is the importance of pilot studies as they can help refine study protocols and overcome confounds.
We hope our experiences designing this user study can help other researchers better design their own studies in the future.

\section{Limitations and Future Work}
\label{section:limitations}
There were five main limitations of our study.
First, our study (designed to be a design probe) lacked a baseline condition where participants complete tasks without interacting with provenance attributes, which would have allowed for a more rigorous comparison of performance between conditions.

Second, our study was conducted in a single session lasting under 90-minutes, which may not fully capture the long-term benefits of visualizing provenance during analysis (i.e., beyond recall-based question-answering tasks); future studies should consider longer durations to validate sustained impact.

Third, while our prototype lets users customize the range of provenance attributes (e.g., darker points can be mapped to smaller or larger values), we disabled this functionality during the study to minimize users' cognitive load. Systematically studying this feature, i.e., if one range leads to more unique data discoveries~\cite{feng2017hindsight} while another leads to more data revisits~\cite{narechania2022lumos} is future work.

Fourth, we currently model focus by equally weighting all interactions but future work can weight differently, e.g., recent interactions more than the older ones~\cite{zhou2021modeling}.

Lastly, we modeled \emph{recency} and \emph{frequency} only based on mouse interactions such as clicks and hovers, which may not be a complete proxy for focus.
In fact, \user{6} expressed some confusion, \emph{``The wording of what `interaction' means was not clear, [I] needed help to understand that.''}
A similar confusion had also come up during our pilot studies, but we tried to address it by clearly explaining to users how provenance is computed, and ensuring that users become acquainted with it during the practice. 
We also posit some of this (recurrent) confusion may be due to participants' surprise upon seeing the system's projection of their provenance not matching their own expectation.
However, we acknowledge that our modeling of provenance from hover interactions with data points, might be \emph{strict} as it does not account for other forms of engagement, like focusing on specific clusters of data points (that remain after applying a filter). Still, we chose it for its practicality and because it likely reflects deliberate user engagement.
Future work may utilize other mechanisms such as user gaze to more accurately model focus (e.g., which attribute(s) and data point(s) is the user actually looking more at).

In general, based on the encouraging findings of this design probe, we next plan to further understand the interplay between data and provenance attributes, particularly how users organically utilize these attributes in high-stakes analysis tasks (unlike a harmless movie selection task as in this study).

\section{Conclusion}
\label{section:conclusion}
In visual data analysis systems, provenance is often logged but only reviewed after analysis, or is hardwired, limiting user access and control. 
In this work, we utilized provenance as an attribute \textit{during} analysis, tracking both \emph{recency} and \emph{frequency} of user interactions with data. 
To evaluate the utility of provenance attributes, we integrated them into a prototype visual data analysis system, \app, which allows users to track and visualize recency and frequency by mapping them to visual encodings (e.g., color or size) and data transformations (sort or filter). 
An exploratory study~with sixteen users found that provenance attributes can help users accurately and confidently review and answer questions about their analysis, often surprising them and facilitating self-reflection.

\bibliographystyle{IEEEtran}
\bibliography{IEEEabrv,template}

\begin{thebibliography}{10}
\providecommand{\url}[1]{#1}
\csname url@samestyle\endcsname
\providecommand{\newblock}{\relax}
\providecommand{\bibinfo}[2]{#2}
\providecommand{\BIBentrySTDinterwordspacing}{\spaceskip=0pt\relax}
\providecommand{\BIBentryALTinterwordstretchfactor}{4}
\providecommand{\BIBentryALTinterwordspacing}{\spaceskip=\fontdimen2\font plus
\BIBentryALTinterwordstretchfactor\fontdimen3\font minus \fontdimen4\font\relax}
\providecommand{\BIBforeignlanguage}[2]{{%
\expandafter\ifx\csname l@#1\endcsname\relax
\typeout{** WARNING: IEEEtran.bst: No hyphenation pattern has been}%
\typeout{** loaded for the language `#1'. Using the pattern for}%
\typeout{** the default language instead.}%
\else
\language=\csname l@#1\endcsname
\fi
#2}}
\providecommand{\BIBdecl}{\relax}
\BIBdecl

\bibitem{north2011analytic}
C.~North, R.~Chang, A.~Endert, W.~Dou, R.~May, B.~Pike, and G.~Fink, ``{Analytic Provenance: Process + Interaction + Insight},'' in \emph{{Extended Abstracts of the CHI Conference on Human Factors in Computing Systems}}.\hskip 1em plus 0.5em minus 0.4em\relax Association for Computing Machinery, 2011, pp. 33--36.

\bibitem{ragan2015characterizing}
E.~D. Ragan, A.~Endert, J.~Sanyal, and J.~Chen, ``{Characterizing Provenance in Visualization and Data Analysis: An Organizational Framework of Provenance Types and Purposes},'' \emph{{IEEE Transactions on Visualization and Computer Graphics}}, vol.~22, no.~1, pp. 31--40, 2016.

\bibitem{narechania2022lumos}
A.~Narechania, A.~Coscia, E.~Wall, and A.~Endert, ``{Lumos: Increasing Awareness of Analytic Behavior during Visual Data Analysis},'' \emph{{IEEE Transactions on Visualization and Computer Graphics}}, vol.~28, no.~1, pp. 1009--1018, 2022.

\bibitem{feng2017hindsight}
M.~Feng, C.~Deng, E.~M. Peck, and L.~Harrison, ``{HindSight: Encouraging Exploration through Direct Encoding of Personal Interaction History},'' \emph{{IEEE Transactions on Visualization and Computer Graphics}}, vol.~23, no.~1, pp. 351--360, 2017.

\bibitem{willett2007scented}
W.~Willett, J.~Heer, and M.~Agrawala, ``{Scented widgets: Improving Navigation Cues with Embedded Visualizations},'' \emph{{IEEE Transactions on Visualization and Computer Graphics}}, vol.~13, no.~6, pp. 1129--1136, 2007.

\bibitem{miller1994magical}
G.~A. Miller, ``{The Magical Number Seven, Plus or Minus Two: Some Limits on Our Capacity for Processing Information},'' \emph{{Psychological Review}}, vol. 101, no.~2, p. 343, 1994.

\bibitem{liu2014effects}
Z.~Liu and J.~Heer, ``{The Effects of Interactive Latency on Exploratory Visual Analysis},'' \emph{{IEEE Transactions on Visualization and Computer Graphics}}, vol.~20, no.~12, pp. 2122--2131, 2014.

\bibitem{stitz2019knowledgepearls}
H.~Stitz, S.~Gratzl, H.~Piringer, T.~Zichner, and M.~Streit, ``{KnowledgePearls: Provenance-Based Visualization Retrieval},'' \emph{IEEE Transactions on Visualization and Computer Graphics}, vol.~25, no.~1, pp. 120--130, 2019.

\bibitem{zhou2021modeling}
\BIBentryALTinterwordspacing
Z.~Zhou, X.~Wen, Y.~Wang, and D.~Gotz, \emph{{Modeling and Leveraging Analytic Focus During Exploratory Visual Analysis}}.\hskip 1em plus 0.5em minus 0.4em\relax New York, NY, USA: Association for Computing Machinery, 2021. [Online]. Available: \url{https://doi.org/10.1145/3411764.3445674}
\BIBentrySTDinterwordspacing

\bibitem{narechania2025provenancewidgets}
A.~Narechania, K.~Odak, M.~El-Assady, and A.~Endert, ``{ProvenanceWidgets: A Library of UI Control Elements to Track and Dynamically Overlay Analytic Provenance},'' \emph{IEEE Transactions on Visualization and Computer Graphics}, vol.~31, no.~1, pp. 1235--1245, 2025.

\bibitem{heer2008graphical}
J.~Heer, J.~Mackinlay, C.~Stolte, and M.~Agrawala, ``{Graphical Histories for Visualization: Supporting Analysis, Communication, and Evaluation},'' \emph{{IEEE Transactions on Visualization and Computer Graphics}}, vol.~14, no.~6, pp. 1189--1196, 2008.

\bibitem{xu2020survey}
K.~Xu, A.~Ottley, C.~Walchshofer, M.~Streit, R.~Chang, and J.~Wenskovitch, ``{Survey on the Analysis of User Interactions and Visualization Provenance},'' \emph{{Computer Graphics Forum}}, vol.~39, no.~3, pp. 757--783, 2020.

\bibitem{perry2009supporting}
J.~Perry, C.~D. Janneck, C.~Umoja, and W.~M. Pottenger, ``{Supporting Cognitive Models of Sensemaking in Analytics Systems},'' \emph{DIMACS}, 2009.

\bibitem{nguyen2016sensemap}
P.~H. Nguyen, K.~Xu, A.~Bardill, B.~Salman, K.~Herd, and B.~W. Wong, ``{SenseMap: Supporting Browser-based Online Sensemaking through Analytic Provenance},'' in \emph{IEEE VAST}, 2016.

\bibitem{madanagopal2019analytic}
K.~Madanagopal, E.~D. Ragan, and P.~Benjamin, ``{Analytic Provenance in Practice: The Role of Provenance in Real-World Visualization and Data Analysis Environments},'' \emph{{IEEE Computer Graphics and Applications}}, vol.~39, no.~6, pp. 30--45, 2019.

\bibitem{bylinskii2017learning}
Z.~Bylinskii, N.~W. Kim, P.~O'Donovan, S.~Alsheikh, S.~Madan, H.~Pfister, F.~Durand, B.~Russell, and A.~Hertzmann, ``{Learning Visual Importance for Graphic Designs and Data Visualizations},'' in \emph{ACM UIST}, 2017.

\bibitem{gomez2012modeling}
S.~Gomez and D.~Laidlaw, ``{Modeling Task Performance for a Crowd of Users from Interaction Histories},'' in \emph{ACM CHI}, 2012.

\bibitem{ceneda2016characterizing}
D.~Ceneda, T.~Gschwandtner, T.~May, S.~Miksch, H.-J. Schulz, M.~Streit, and C.~Tominski, ``{Characterizing Guidance in Visual Analytics},'' \emph{{IEEE Transactions on Visualization and Computer Graphics}}, vol.~23, no.~1, pp. 111--120, 2017.

\bibitem{walch2019lightguider}
A.~Walch, M.~Schwärzler, C.~Luksch, E.~Eisemann, and T.~Gschwandtner, ``{LightGuider: Guiding Interactive Lighting Design using Suggestions, Provenance, and Quality Visualization},'' \emph{{IEEE Transactions on Visualization and Computer Graphics}}, vol.~26, no.~1, pp. 569--578, 2020.

\bibitem{endert2012semanticinteraction}
A.~Endert, P.~Fiaux, and C.~North, ``{Semantic Interaction for Visual Text Analytics},'' in \emph{{Proceedings of the SIGCHI Conference on Human Factors in Computing Systems}}, 2012, pp. 473--482.

\bibitem{bavoil2005vistrails}
L.~Bavoil, S.~P. Callahan, P.~J. Crossno, J.~Freire, C.~E. Scheidegger, C.~T. Silva, and H.~T. Vo, ``{Vistrails: Enabling Interactive Multiple-View Visualizations},'' in \emph{{VIS 05. IEEE Visualization, 2005.}}\hskip 1em plus 0.5em minus 0.4em\relax IEEE, 2005, pp. 135--142.

\bibitem{shrinivasan2009connecting}
Y.~B. Shrinivasan, D.~Gotz, and J.~Lu, ``{Connecting the Dots in Visual Analysis},'' in \emph{IEEE VAST}, 2009.

\bibitem{chen2010click2annotate}
Y.~Chen, S.~Barlowe, and J.~Yang, ``{Click2Annotate: Automated Insight Externalization with Rich Semantics},'' in \emph{2010 IEEE Symposium on Visual Analytics Science and Technology}.\hskip 1em plus 0.5em minus 0.4em\relax IEEE, 2010, pp. 155--162.

\bibitem{gratzl2016visual}
S.~Gratzl, A.~Lex, N.~Gehlenborg, N.~Cosgrove, and M.~Streit, ``{From Visual Exploration to Storytelling and Back Again},'' in \emph{Computer Graphics Forum}, 2016.

\bibitem{wall2022lrg}
E.~Wall, A.~Narechania, A.~Coscia, J.~Paden, and A.~Endert, ``{Left, Right, and Gender: Exploring Interaction Traces to Mitigate Human Biases},'' \emph{{IEEE Transactions on Visualization and Computer Graphics}}, vol.~28, no.~1, pp. 966--975, 2022.

\bibitem{block2023influence}
J.~E. Block, S.~Esmaeili, E.~D. Ragan, J.~R. Goodall, and G.~D. Richardson, ``{The Influence of Visual Provenance Representations on Strategies in a Collaborative Hand-off Data Analysis Scenario},'' \emph{{IEEE Transactions on Visualization and Computer Graphics}}, vol.~29, no.~1, pp. 1113--1123, 2023.

\bibitem{footstepsvscode}
A.~Wattenberger, ``{Footsteps for VS Code},'' \url{https://marketplace.visualstudio.com/items?itemName=Wattenberger.footsteps}, 2021, {Accessed: June 15, 2024}.

\bibitem{gadhave2024persist}
K.~Gadhave, Z.~Cutler, and A.~Lex, ``{Persist: Persistent and Reusable Interactions in Computational Notebooks},'' in \emph{Computer Graphics Forum}, 2024.

\bibitem{eckelt2024loops}
\BIBentryALTinterwordspacing
K.~Eckelt, K.~Gadhave, A.~Lex, and M.~Streit, ``{Loops: Leveraging Provenance and Visualization to Support Exploratory Data Analysis in Notebooks},'' \emph{{OSF Preprint}}, 2023. [Online]. Available: \url{https://doi.org/10.31219/osf.io/79eyn}
\BIBentrySTDinterwordspacing

\bibitem{revisit2023ding}
Y.~Ding, J.~Wilburn, H.~Shrestha, A.~Ndlovu, K.~Gadhave, C.~Nobre, A.~Lex, and L.~Harrison, ``{reVISit: Supporting Scalable Evaluation of Interactive Visualizations},'' in \emph{IEEE VIS}, 2023.

\bibitem{ellkvist2008using}
T.~Ellkvist, D.~Koop, E.~W. Anderson, J.~Freire, and C.~Silva, ``{Using Provenance to Support Real-Time Collaborative Design of Workflows},'' in \emph{Provenance and Annotation of Data and Processes}, 2008.

\bibitem{sarvghad2015exploiting}
A.~Sarvghad and M.~Tory, ``{Exploiting Analysis History to Support Collaborative Data Analysis},'' in \emph{{Proceedings of the 41st Graphics Interface Conference}}, 2015, pp. 123--130.

\bibitem{badam2017supporting}
S.~K. Badam, Z.~Zeng, E.~Wall, A.~Endert, and N.~Elmqvist, ``{Supporting Team-First Visual Analytics through Group Activity Representations.}'' in \emph{Graphics Interface}, 2017.

\bibitem{googleanalytics}
``{Google Analytics},'' \url{https://marketingplatform.google.com/about/analytics}, {Accessed: June 15, 2024}.

\bibitem{hotjar}
``{Hotjar},'' \url{https://www.hotjar.com}, {Accessed: June 15, 2024}.

\bibitem{drachen2015behavioral}
A.~Drachen, ``{Behavioral Telemetry in Games User Research},'' \emph{{Game User Experience Evaluation}}, pp. 135--165, 2015.

\bibitem{kohwalter2017capturing}
T.~C. Kohwalter, L.~G.~P. Murta, and E.~W.~G. Clua, ``{Capturing Game Telemetry with Provenance},'' in \emph{{2017 16th Brazilian Symposium on Computer Games and Digital Entertainment (SBGames)}}.\hskip 1em plus 0.5em minus 0.4em\relax IEEE, 2017, pp. 66--75.

\bibitem{callahan2006vistrails}
S.~P. Callahan, J.~Freire, E.~Santos, C.~E. Scheidegger, C.~T. Silva, and H.~T. Vo, ``{VisTrails: Visualization meets Data Management},'' in \emph{{Proceedings of the 2006 ACM SIGMOD international conference on Management of data}}, 2006, pp. 745--747.

\bibitem{aigner2013evalbench}
W.~Aigner, S.~Hoffmann, and A.~Rind, ``{EvalBench: A Software Library for Visualization Evaluation},'' in \emph{{Computer Graphics Forum}}, vol.~32, no. 3pt1.\hskip 1em plus 0.5em minus 0.4em\relax Wiley Online Library, 2013, pp. 41--50.

\bibitem{okoe2015graphunit}
M.~Okoe and R.~Jianu, ``{GraphUnit: Evaluating Interactive Graph Visualizations Using Crowdsourcing},'' in \emph{{Computer Graphics Forum}}, vol.~34, no.~3.\hskip 1em plus 0.5em minus 0.4em\relax Wiley Online Library, 2015, pp. 451--460.

\bibitem{cutler2020trrack}
Z.~{Cutler}, K.~{Gadhave}, and A.~{Lex}, ``{Trrack: A Library for Provenance-Tracking in Web-Based Visualizations},'' in \emph{{2020 IEEE Visualization Conference (VIS)}}, 2020, pp. 116--120.

\bibitem{arroyo2006usability}
E.~Arroyo, T.~Selker, and W.~Wei, ``{Usability tool for analysis of web designs using mouse tracks},'' in \emph{{CHI'06 extended abstracts on Human factors in computing systems}}, 2006, pp. 484--489.

\bibitem{nielsen2010eyetracking}
J.~Nielsen and K.~Pernice, \emph{{Eyetracking web usability}}.\hskip 1em plus 0.5em minus 0.4em\relax New Riders, 2010.

\bibitem{baudisch2006phosphor}
P.~Baudisch, D.~Tan, M.~Collomb, D.~Robbins, K.~Hinckley, M.~Agrawala, S.~Zhao, and G.~Ramos, ``{Phosphor: Explaining Transitions in the User Interface Using Afterglow Effects},'' in \emph{{Proceedings of the 19th annual ACM symposium on User Interface Software and Technology}}, 2006, pp. 169--178.

\bibitem{hill2003awareness}
J.~Hill and C.~Gutwin, ``{Awareness Support in a Groupware Widget Toolkit},'' in \emph{{Proceedings of the 2003 ACM International Conference on Supporting Group Work}}, 2003, pp. 258--267.

\bibitem{angelini2020crosswidgets}
M.~Angelini, G.~Blasilli, S.~Lenti, A.~Palleschi, and G.~Santucci, ``{CrossWidgets: Enhancing Complex Data Selections through Modular Multi Attribute Selectors},'' in \emph{Proceedings of the International Conference on Advanced Visual Interfaces}, 2020, pp. 1--9.

\bibitem{gutwin2002traces}
C.~Gutwin, ``{Traces: Visualizing the immediate past to support group interaction},'' in \emph{{Graphics interface}}.\hskip 1em plus 0.5em minus 0.4em\relax Citeseer, 2002, pp. 43--50.

\bibitem{boy2015storytelling}
J.~Boy, F.~Detienne, and J.-D. Fekete, ``{Storytelling in information visualizations: Does it engage users to explore data?}'' in \emph{ACM CHI}, 2015.

\bibitem{feng2019patterns}
M.~Feng, E.~Peck, and L.~Harrison, ``{Patterns and Pace: Quantifying Diverse Exploration Behavior with Visualizations on the Web},'' \emph{{IEEE Transactions on Visualization and Computer Graphics}}, vol.~25, no.~1, pp. 501--511, 2019.

\bibitem{ottley2019follow}
A.~Ottley, R.~Garnett, and R.~Wan, ``{Follow The Clicks: Learning and Anticipating Mouse Interactions During Exploratory Data Analysis},'' in \emph{{Computer Graphics Forum}}, vol.~38, no.~3.\hskip 1em plus 0.5em minus 0.4em\relax Wiley Online Library, 2019, pp. 41--52.

\bibitem{gotz2016adaptive}
D.~Gotz, S.~Sun, and N.~Cao, ``{Adaptive Contextualization: Combating Bias During High-Dimensional Visualization and Data Selection},'' in \emph{{Proceedings of the 21st International Conference on Intelligent User Interfaces}}, 2016, pp. 85--95.

\bibitem{wall2017warning}
E.~Wall, L.~M. Blaha, L.~Franklin, and A.~Endert, ``{Warning, Bias May Occur: A Proposed Approach to Detecting Cognitive Bias in Interactive Visual Analytics},'' in \emph{{2017 IEEE Conference on Visual Analytics Science and Technology (VAST)}}, 2017, pp. 104--115.

\bibitem{skopik2005improving}
A.~Skopik and C.~Gutwin, ``{Improving Revisitation in Fisheye Views with Visit Wear},'' in \emph{{Proceedings of the SIGCHI conference on Human Factors in Computing Systems}}, 2005, pp. 771--780.

\bibitem{kaasten2002people}
S.~Kaasten, S.~Greenberg, and C.~Edwards, ``{How People Recognise Previously Seen Web Pages from Titles, URLs and Thumbnails},'' in \emph{HCI}, 2002.

\bibitem{hill1992edit}
W.~C. Hill, J.~D. Hollan, D.~Wroblewski, and T.~McCandless, ``{Edit Wear and Read Wear},'' in \emph{{Proceedings of the SIGCHI Conference on Human Factors in Computing Systems}}, 1992, pp. 3--9.

\bibitem{alexander2009revisiting}
J.~Alexander, A.~Cockburn, S.~Fitchett, C.~Gutwin, and S.~Greenberg, ``{Revisiting Read Wear: Analysis, Design, and Evaluation of a Footprints Scrollbar},'' in \emph{{Proceedings of the SIGCHI Conference on Human Factors in Computing Systems}}, 2009, pp. 1665--1674.

\bibitem{oliveira2017framework}
W.~Oliveira, L.~M. Ambr{\'o}sio, R.~Braga, V.~Str{\"o}ele, J.~M. David, and F.~Campos, ``{A Framework for Provenance Analysis and Visualization},'' \emph{Procedia Computer Science}, 2017.

\bibitem{kohwalter2016prov}
T.~Kohwalter, T.~Oliveira, J.~Freire, E.~Clua, and L.~Murta, ``{Prov Viewer: A graph-based visualization tool for interactive exploration of provenance data},'' in \emph{IPAW}, 2016.

\bibitem{chen2012visualization}
P.~Chen, B.~Plale, Y.-W. Cheah, D.~Ghoshal, S.~Jensen, and Y.~Luo, ``{Visualization of Network Data Provenance},'' in \emph{HiPC}, 2012.

\bibitem{shneiderman1983direct}
B.~Shneiderman, ``{Direct manipulation: A step beyond programming languages},'' \emph{Computer}, 1983.

\bibitem{wolter2009direct}
M.~Wolter, B.~Hentschel, I.~Tedjo-Palczynski, and T.~Kuhlen, ``{A direct manipulation interface for time navigation in scientific visualizations},'' in \emph{IEEE 3DUI}, 2009.

\bibitem{kondo2014dimpvis}
B.~Kondo and C.~Collins, ``{Dimpvis: Exploring time-varying information visualizations by direct manipulation},'' \emph{{IEEE Transactions on Visualization and Computer Graphics}}, 2014.

\bibitem{lex2014upset}
A.~Lex, N.~Gehlenborg, H.~Strobelt, R.~Vuillemot, and H.~Pfister, ``{UpSet: visualization of intersecting sets},'' \emph{{IEEE Transactions on Visualization and Computer Graphics}}, 2014.

\bibitem{tableau}
\BIBentryALTinterwordspacing
Tableau, ``{Tableau},'' 2022. [Online]. Available: \url{https://www.tableau.com/learn/get-started/prep}
\BIBentrySTDinterwordspacing

\bibitem{kandel2011wrangler}
S.~Kandel, A.~Paepcke, J.~Hellerstein, and J.~Heer, ``{Wrangler: Interactive Visual Specification of Data Transformation Scripts},'' in \emph{{Proceedings of the SIGCHI Conference on Human Factors in Computing Systems}}, 2011, pp. 3363--3372.

\bibitem{narechania2023datapilot}
A.~Narechania, F.~Du, A.~R. Sinha, R.~Rossi, J.~Hoffswell, S.~Guo, E.~Koh, S.~B. Navathe, and A.~Endert, ``{DataPilot: Utilizing Quality and Usage Information for Subset Selection during Visual Data Preparation},'' in \emph{{Proceedings of the 2023 CHI Conference on Human Factors in Computing Systems}}, 2023.

\bibitem{narechania2023datacockpit}
A.~Narechania, S.~Chakraborty, S.~Agarwal, A.~R. Sinha, R.~A. Rossi, F.~Du, J.~Hoffswell, S.~Guo, E.~Koh, A.~Endert, and S.~Navathe, ``{DataCockpit: A Toolkit for Data Lake Navigation and Monitoring Utilizing Quality and Usage Information},'' in \emph{{2023 IEEE International Conference on Big Data (BigData)}}, 2023, pp. 5305--5310.

\bibitem{satyanarayan2016vega}
A.~Satyanarayan, D.~Moritz, K.~Wongsuphasawat, and J.~Heer, ``{Vega-Lite: A Grammar of Interactive Graphics},'' \emph{{IEEE Transactions on Visualization and Computer Graphics}}, vol.~23, no.~1, pp. 341--350, 2017.

\bibitem{franconeri2021scienceviscom}
\BIBentryALTinterwordspacing
S.~L. Franconeri, L.~M. Padilla, P.~Shah, J.~M. Zacks, and J.~Hullman, ``{The Science of Visual Data Communication: What Works},'' \emph{Psychological Science in the Public Interest}, vol.~22, no.~3, pp. 110--161, 2021. [Online]. Available: \url{https://doi.org/10.1177/15291006211051956}
\BIBentrySTDinterwordspacing

\bibitem{nobre2021revisit}
C.~Nobre, D.~Wootton, Z.~Cutler, L.~Harrison, H.~Pfister, and A.~Lex, ``{reVISit: Looking under the hood of interactive visualization studies},'' in \emph{{Proceedings of the 2021 CHI Conference on Human Factors in Computing Systems}}, 2021, pp. 1--13.

\bibitem{brooke1996sus}
J.~Brooke, ``{SUS: A Quick and Dirty Usability Scale},'' \emph{{Usability Evaluation in Industry}}, vol. 189, no. 194, pp. 4--7, 1996.

\bibitem{nickerson1998confirmation}
R.~S. Nickerson, ``{Confirmation Bias: A Ubiquitous Phenomenon in Many Guises},'' \emph{{Review of General Psychology}}, 1998.

\bibitem{atkinson1968human}
R.~C. Atkinson and R.~M. Shiffrin, ``{Human Memory: A Proposed System and its Control Processes},'' in \emph{{Psychology of Learning and Motivation}}, 1968.

\bibitem{tversky1973availability}
A.~Tversky and D.~Kahneman, ``{Availability: A Heuristic for Judging Frequency and Probability},'' \emph{{Cognitive Psychology}}, 1973.

\bibitem{sperrle2021co}
F.~Sperrle, A.~Jeitler, J.~Bernard, D.~Keim, and M.~El-Assady, ``{Co-Adaptive Visual Data Analysis and Guidance Processes},'' \emph{Computers \& Graphics}, vol. 100, pp. 93--105, 2021.

\end{thebibliography}

\vspace{-2.5em}
\begin{IEEEbiography}[{\includegraphics[width=1in,height=1.25in,clip,keepaspectratio]{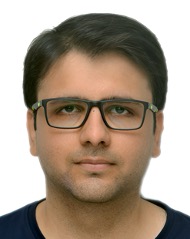}}]{Arpit Narechania}
is an assistant professor in the Department of Computer Science and Engineering at The Hong Kong University of Science and Technology (HKUST). He directs the DataVisards group, which explores novel data experiences for human-AI collaboration for visual analytics.
\end{IEEEbiography}

\vspace{-3.4em}
\begin{IEEEbiography}[{\includegraphics[width=1in,height=1.25in,clip,keepaspectratio]{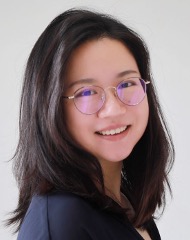}}]{Shunan Guo}
is a research scientist at Adobe. Her research interest lies in the area of information visualization and visual analytics, with a focus on designing and developing visual analytics prototypes that communicate insights derived from complex data models.
\end{IEEEbiography}

\vspace{-3.4em}
\begin{IEEEbiography}[{\includegraphics[width=1in,height=1.25in,clip,keepaspectratio]{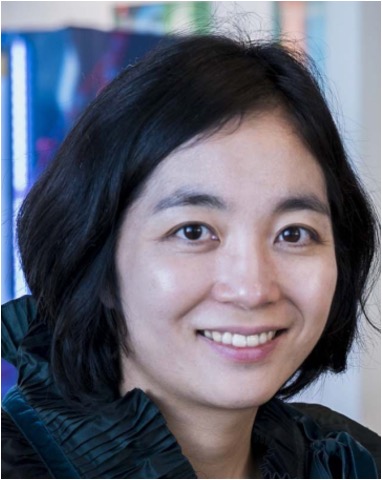}}]{Eunyee Koh}
is a principal scientist at Adobe. Her research interest lies at the intersection of human-computer interaction, AI/machine learning, and data science. She is passionate about how to personalize information for people.
\end{IEEEbiography}

\vspace{-3.4em}
\begin{IEEEbiography}[{\includegraphics[width=1in,height=1.25in,clip,keepaspectratio]{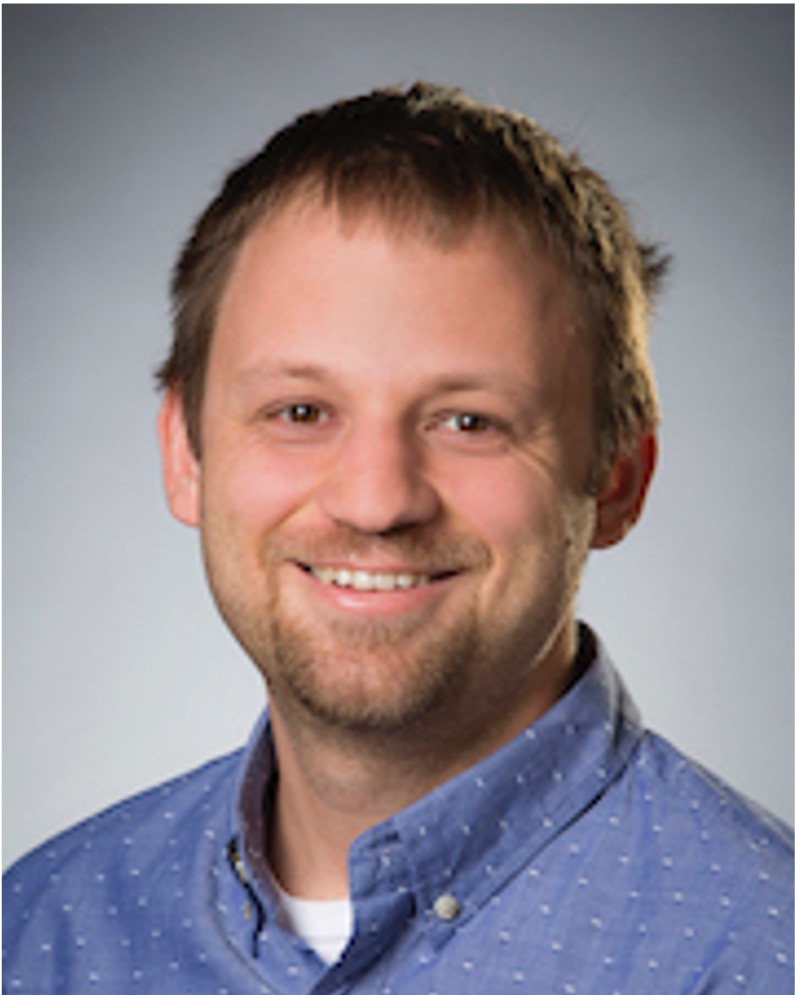}}]{Alex Endert}
is an associate professor in the School of Interactive Computing, Georgia Tech. He directs the Visual Analytics Lab, which explores novel user interaction techniques for visual analytics.
\end{IEEEbiography}

\vspace{-3.4em}
\begin{IEEEbiography}[{\includegraphics[width=1in,height=1.25in,clip,keepaspectratio]{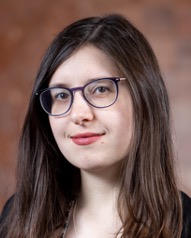}}]{Jane Hoffswell}
is a research scientist at Adobe, specializing in visualization and human-computer interaction. Her research explores the design of responsive visualizations, interactive systems, and data-driven visual storytelling.
\end{IEEEbiography}

\end{document}